\documentstyle[12pt]{article}
\newcommand{\be}{\begin{equation}}
\newcommand{\ee}{\end{equation}}
\newcommand{\k}{{p}}
\newcommand{\ii}{{\it i}}
\newcommand{\f}{\varphi}
\newcommand{\iii}{\int_{0}^{\infty}}
\newcommand{\dd}{{Phys.Rev.\ }}
\newcommand{\prl}{{Phys.Rev.Lett.\ }}
\newcommand{\pl}{{Phys.Lett.\ }}
\newcommand{\np}{{Nucl.Phys.\ }}
\begin{document}
\begin{titlepage}
\begin{flushright}
Z\"urich University Preprint\\
ZU-TH 1/94\\
January 1994\\

\end{flushright}
\vspace{10 mm}
\begin{center}
\huge
Fermion number non-conservation and gravity
\vspace{5 mm}
\end{center}
\begin{center}
{\bf  Mikhail S. Volkov}\footnote{On  leave  from  Physical-Technical
Institute of the  Academy  of  Sciences  of  Russia,  Kazan  420029,
Russia}

\vspace{5 mm}
Institut    f\"ur    Theoretische    Physik    der    Universit\"at
Z\"urich-Irchel, Winterthurerstrasse 190, CH-8057 Z\"urich,
Switzerland

e-mail:\ volkov@physik.unizh.ch
\end{center}
\vspace{10 mm}
\begin{center}
{\large Abstract}
\end{center}
\vspace{5 mm}

It is shown that in the Einstein-Yang-Mills (EYM) theory, as well as
in the pure flat space Yang-Mills (YM) theory, there  always  exists
an  opportunity  to  pass  over  the  potential  barrier  separating
homotopically distinct vacuum sectors, because  the  barrier  height
may  be  arbitrarily  small.   However,  at  low  energies  all  the
overbarrier   histories   are   suppressed   by   the    destructive
interference. In the pure YM theory the situation remains  the  same
for any energies. In the EYM theory on  the  other  hand,  when  the
energy is large and exceeds the ground state EYM sphaleron mass, the
constructive interference occurs instead. This  means  that  in  the
extreme high energy limit the exponential suppression of the fermion
number violation in  pure YM theory is removed due to  gravitational
effects.
\end{titlepage}
\newpage

\section{Introduction}

     The discovery of the vacuum   periodicity   in   a   Yang-Mills
(YM) theory \cite{1} gave rise to the  theoretical  description   of
baryon number  non-conservation \cite{2}  caused  by  the  anomalous
violation    of  chirality  \cite{3}.  The  standard   approach   to
understanding   this   phenomenon   uses    semiclassical    barrier
penetration between topologically inequivalent  vacua,  where    the
tunneling  solutions  are   Euclidean   instantons  \cite{4}.    The
amplitude  of  such a penetration at zero energy is of the order  of
$exp(-S^{Eucl})$, which is incredibly small  because  of  the  large
value  of  the  instanton  action $S^{Eucl}$. In a  pure  Yang-Mills
theory  one  can  not  expect the amplitude  to  become  large  with
growing  energy,  because  the   theory   does   not   contain   any
dimensional parameters to set the energy scale.

     When scalar fields are involved, as in Yang-Mills-Higgs   (YMH)
theory, a natural energy scale arises, and one may think  that   the
transition amplitude will not be small if the energy is of the order
of some threshold value. Indeed,  the  existence  of  the  sphaleron
solution in the electroweak YMH theory  \cite{5}  implies  that  the
fermion number  violating  processes may become unsuppressed if  the
energy (or  temperature)  is  of  the order of  the  sphaleron  mass
\cite{6,7}. The idea is that, instead of  tunnel  penetration  under
the barrier, the field configuration can pass   over   it   and  the
value of the sphaleron energy is the {\it minimal} barrier height.

     If the energy is very high, then the symmetry is restored.  The
YMH sphalerons then dissociate, and  the  fermion  number  violating
processes   again   become  suppressed  \cite{8}.  However,  another
universal  energy  scale   arises   naturally   at  extremely   high
energies, namely, the gravitational Planck  scale. The existence  of
this  new  scale  leads  one  to  expect  that  the  fermion  number
violation in the YM theory will become significant when  the  energy
is  comparable with Planck's energy. This expectation  is  supported
by   the   discovery    of   particle-like    solutions    in    the
Einstein-Yang-Mills   (EYM)   theory   \cite{9},  which  have   been
interpreted as sphalerons  \cite{10}.   One  might  think  that  the
typical mass of these EYM  sphalerons, similarly to the situation in
the YMH theory, determines the transition threshold.

It turns out however, that, contrary to the  situation  in  the  YMH
theory, none of the EYM sphalerons  relate  to  the  highest  energy
point on a minimal-energy path connecting  inequivalent  EYM  vacua,
for the potential barrier between distinct vacua in the  EYM  theory
can be arbitrarily low \cite{10}. Indeed, let us consider first  the
flat space YM   theory  and   define   a  one-parameter   continuous
family  of   static   field   configurations,  $A(\vec{x},\lambda)$,
such that when the parameter, $\lambda$,  goes from zero to   unity,
the  field  $A(\vec{x},\lambda)$   interpolates  between  two  vacua
with distinct winding numbers. We will  call such a family of fields
a   vacuum-to-vacuum  (VTV)  path.   The   simplest    choice     is
\cite{1}  $A(\vec{x},  \lambda)  =\lambda  A^{(1)}(\vec{x})$,  where
$A^{(1)}(\vec{x})$ is a pure gauge with unit  winding  number.   The
energy for  this  family,  $U(\lambda)$,  has  the  typical  barrier
behaviour  as it vanishes for the vacuum values, $\lambda =0,1$, and
reaches a maximum in the middle at $\lambda = 1/2$. This allows  one
to say that distinct  vacua are   divided   by   a   barrier.   Next
note,  that  continuous   scaling deformation   of    the    family,
$A(\vec{x},\lambda)\rightarrow  A(\beta\vec{x},\lambda)$,     alters
the     energy     as      follows:   $U(\lambda)\rightarrow   \beta
U(\lambda)$,   and   hence   the barrier height, $\beta U(1/2)$, can
be made arbitrarily low  by a proper choice of the  scaling   factor
$\beta$.  It  is  clear  that the  inclusion  of  gravity  can  only
reduce the energy and  hence  does   not  change  the  situation  (a
consistent way to  determine  of   the  gravitating  energy  demands
solving the initial value  constraints; see below).

     Thus, the EYM sphalerons, having finite mass, are not   related
to the {\it minimal} barrier height. One may wonder then what  their
physical  meaning  is.  Indeed, as the   barrier   height   may   be
arbitrarily small, an overbarrier passage is possible at   any   low
energy and not only at the   (ultra-high)   sphaleron   energy.   In
other words, there is no need to climb  the  mountain  in  order  to
reach the neighbouring valley as it is possible  to  go  around  it.
This  certainly  obscures  the  threshold  interpretation   of   the
sphalerons.

     However, there exists also a counter example to this objection.
Indeed, in the flat space limit, where the  overbarrier  passage  is
also possible at  any  non-zero  energy,  the  transition  amplitude
nevertheless always remains small (in the {\it pure} YM theory),  as
all fermion-violating Green functions  contain the small   instanton
factor, regardless of the value of the energy  \cite{7}.  Therefore,
one may infer that the possibility to pass over  the  barrier   does
not  yet  guarantee in itself an enhancement of the transition rate.
To understand why this may happen one has to consider not  only  one
path  interpolating  between  distinct vacua, but  all  such  paths,
which   corresponds   to   the  inclusion   of  all   the   possible
transition histories. The contribution of a   single   path  to  the
transition amplitude  may  not  be  small  provided that  this  path
lies entirely in  the  classically  allowed  region,  that  is,  the
energy exceeds the barrier height for the given  path  --  we  shall
call  such a path an {\it overbarrier path}. However, when one takes
the sum over all histories, it may happen that the contributions  of
different paths cancel each other. The latter occurs when the set of
all  paths  does  not include a path  whose  contribution  into  the
transition  amplitude is stationary with respect to small variations
of the path itself. Such a situation  is  encountered  in  the  flat
space  pure  YM  theory,  where  all  overbarrier   histories    are
strongly  suppressed  by  the  destructive   interference,
regardless of the value of the energy, and a stationary phase  path,
if any  exists,  may only be {\it underbarrier}, so that the   total
transition amplitude is always small.

     One may infer from this example  that for the EYM   theory  all
the fermion number violating transitions should also be   suppressed
in  the  low  energy  limit,  despite  the    existence    of    the
overbarrier paths, for gravitation can  not  essentially  alter  the
dynamics  at low  energies,  and   the  theory    should    resemble
the  flat  space  YM   theory   in   this    limit.  Therefore,  the
vanishing of the greatest  lower  bound of the barrier height may be
not  so  fatal  for  the  threshold  interpretation   of   the   EYM
sphalerons. Indeed, as we  will see below, the  resemblance  between
the YM and the EYM theories only holds at low energies. In the  high
energy  limit, the situation  changes.  Firstly,  at  high  energies
the   set   of   the overbarrier paths in the   EYM   case   becomes
larger  in  comparison  with   that  appearing  in  the  flat  space
theory, because  the gravitational   binding  tends  to  reduce  the
barrier height. In addition, and this is the crucial point, when the
energy exceeds the ground state sphaleron  mass,  gravity introduces
a stationary phase overbarrier history.  This leads us to conclusion
that the transition becomes  unsuppressed  at  high   energies   and
restores the initial threshold interpretation of the sphalerons.

     The  main  goal  of  this  paper  is  to   argue,   using   the
quasi-classical arguments, that  the  suppression   of  the  fermion
number violating transitions in the EYM  theory is removed when  the
energy is larger  than  the  ground  state  sphaleron mass (see also
\cite{main}). The theoretical tool relevant  for  our  purposes  has
been developed by Bitar and Chang in their  real  time  analysis  of
tunneling transitions in the flat space  YM  theory  \cite{11}.  The
essence of this method is  to introduce firstly a VTV  path  in  the
configuration function space of a theory. Then one assumes that  the
transition between distinct vacuum sectors of  the  theory  proceeds
along  this  single  VTV  path.  This  reduces  the  problem  to   a
one-dimensional quantum-mechanical task and  allows  one  to   write
down  immediately the corresponding WKB amplitude. The last step  is
to take the sum over all VTV paths.

We apply Bitar and Chang's technique  to represent  (in  the  lowest
order  WKB  approximation)  the  amplitude  of  the  fermion  number
violating transition in the EYM theory at non-zero energy (a similar
idea was exploited also within the context of the  YMH  theory;  see
Ref.\cite{ringwald}). Following this  line,  we introduce a  set  of
VTV  paths  connecting  homotopically  distinct  vacua  in  the  EYM
configuration function space (see Eqs.(\ref{22}),(\ref{33})  below).
Our  basic  approximation  is  the  assumption  that   the   quantum
transition between distinct vacuum sectors occurs only  along  these
paths.  We  estimate  the  partial  WKB  amplitudes  for  the  paths
introduced and specify those paths whose amplitudes are  not  small.
Such paths lie entirely in the classically allowed region  and  they
relate to overbarrier passages. Remaining paths  pass  through  the
classically  forbidden  region  and  we  exclude   them   from   our
consideration because the corresponding partial amplitudes are small
due to the tunneling suppression. The total transition amplitude  is
the sum over all partial amplitudes which can be estimated by making
use  of  the  stationary  phase  approximation.  We  show  that  the
overbarrier paths always exist, that is,  the  overbarrier  passages
are possible at any non-zero energy. However,  when  the  energy  is
small,  the  set  of  the  overbarrier  paths  does  not  contain  a
stationary phase path. This means, that the sum over all overbarrier
histories, being evaluated via the stationary  phase  approximation,
will be small, as the stationary point is absent, so the overbarrier
passages are suppressed by the destructive interference.  On  the  other
hand, when  the  energy  is  large  and  exceeds  the  ground  state
sphaleron mass, a stationary phase path appears on the  set  of  the
overbarrier paths. This allows us to conclude that, when the  energy
is large, the total transition amplitude in the EYM  theory  is  not
small  (because  the  stationary  point  is  present).   Thus,   the
suppression of the fermion number violation in  the  EYM  theory  is
removed when the energy is of the order of the EYM sphaleron mass.

      The rest of the paper is organized as follows.

     In Sec.II we investigate the homotopical  classification of the
vacua arising  in  the  SU(2)  EYM  theory  and  estimate  instanton
contributions to the fermion number violation rate.  In  Sec.III  we
introduce a set of vacuum-to-vacuum paths in  the EYM  configuration
function space such that the time evolution  along these paths leads
to a change of the topological winding number. In  Sec.IV  we  apply
Bitar and Chang's technique to obtain the WKB transition  amplitude.
To select the overbarrier paths we study in Sec.V the structure   of
the potential barrier dividing vacua  in  the  EYM  theory.  In  the
vicinity of the $n$-th sphaleron solution, we find that locally, the
potential barrier surface is a saddle with one transversal  and  $n$
longitudinal negative directions. Exploiting   local  properties  of
the potential barrier surface we show in  Sec.VI the existence    of
the (approximative)  overbarrier  stationary  phase  path  when  the
energy  exceeds  the ground state sphaleron  mass.  Some  concluding
remarks are  made  in  Sec.VII.  There  are  also  three  appendices
included. Appendix A contains an explanation of our  choice  of  the
gravitational action, and also details of  the  computation  of  the
action for the spherically symmetric case. The Appendices  B  and  C
include  various  formulae  relevant  for  the  description  of  the
spherically symmetric  EYM  fields  and an explicit  procedure   for
the first and second  variation of the ADM energy functional.

     Our  sign  conventions used  are those of  Landau  \&  Lifshitz
\cite{ll}.


\section{The EYM vacua and the instanton estimates}

The starting point of  our  considerations  is  the  notion  of  the
topological vacua of the EYM fields. The topological vacua of the YM
field have been first introduced in Ref.\cite{1} within the  context
of the flat space theory. It turns out  that  inclusion  of  gravity
modifies the analysis in \cite{1} only slightly, as long as one does
not take into account the topological effects of  the  gravitational
field itself.

Consider the action of the EYM theory with the $SU(2)$ gauge group
\be
S_{EYM} = S_{G}+S_{YM},                                   \label{1}
\ee
with the gravitational part
\be
S_{G}=-{1\over 16\pi G}\int R\sqrt{-g}d^{4}x -
{1\over 16\pi  G}\oint_{\Sigma}
(g^{\mu\alpha}\Gamma^{\beta}_{\alpha\beta}-
g^{\alpha\beta}\Gamma^{\mu}_{\alpha\beta})d\Sigma_{\mu},
\ee                                                     \label{2}
and the Yang-Mills contribution
\be
S_{YM}=-{1\over 2g^{2}}\int trF_{\mu\nu}F^{\mu\nu}\sqrt{-g}d^{4}x.
                                                        \label{3}
\ee
Here $G$ is Newton's constant, $g$ is the gauge  coupling  constant,
$F_{\mu\nu}=\partial_{\nu}A_{\mu}-\partial_{\nu}A_{\mu}-        {\it
i}[A_{\mu},A_{\nu}]$ is the  matrix  valued  gauge   field   tensor,
$A_{\mu}=A_{\mu}^{a}\tau^{a}/2$, and $\tau^{a}\ (a=1,2,3)$  are  the
Pauli matrices. We use the  gravitational  action,  which  does  not
contain second derivatives of the metric; for explanations
of our choice see Appendix A.

The  action  (\ref{1})  has  many   stationary   points,   including
solutions  with non-trivial space-time topology such as black  holes
\cite{12}. Consider a sector with spacetime manifolds  carrying  the
trivial $R^{4}$  topology.  Define  vacua  in  this  sector  as  the
stationary points  with  zero  ADM  energy.   As  follows  from  the
positive energy theorem \cite{16}, such  vacua  must   have  a  flat
metric and hence the YM field is a  pure  gauge.  In  the  $A_{0}=0$
gauge one can represent the vacuum fields as
\be
A_{j}(\vec{x})={\it  i}U\partial_{j}U^{-1}, \ \ \
g_{\mu\nu}(\vec{x})=\eta_{\mu\nu},
  \label{6}        \ee
where $\eta_{\mu\nu}$ is the flat space  metric,  $U(\vec{x})$  is a
$SU(2)$ valued function, and  spatial  index  $j$  runs  over  1,2,3
$(\vec{x} \equiv x^{j})$.

Such vacua are to be the boundary conditions for the path  integral.
It seems that not all of them are physically  important,  but   only
those satisfying some  additional  topological   conditions,   which
have the origin in instanton physics. Following  the  standard  line
\cite{1}, we pass to the Euclidean  sector  and  use  the  instanton
arguments to clarify the structure  of  the  EYM  vacua. One  should
note that restricting ourselves only to spacetime manifolds with the
$R^{4}$ topology, we  exclude  from  consideration  all  non-trivial
(i.e.   carrying   non-zero   topological   charges)   gravitational
instantons and vacua. A complete investigation  of  quantum  gravity
effects   requires  inclusion  of  instantons   and    vacua    with
non-trivial  topological   indices,   both   for   the   gauge   and
gravitational fields. This  lies, however,  beyond  the   scope   of
our  present analysis (description  of the non-trivial gravitational
vacua has  been  done  in  Ref.\cite{17}).

Consider  an  instanton  interpolating  between   two   EYM   vacua.
The finiteness of the instanton action  implies  the  finiteness  of
both the matter and the gravitational parts of the action as both of
them are  positive. Indeed,  the  matter  action  is  positive.  The
gravitational  Euclidean action  may  be  arbitrary,   however   the
positive  action  conjecture implies that it should be  positive  on
shell,  where  the  condition $R=0$ holds  \cite{perry}.  Finiteness
of the matter part of the action  implies  that the YM  field  tends
to  a  pure  gauge,  ${\it  i}U\partial_{j}U^{-1}$,   at   Euclidean
infinity,    $S^{3}_{Eucl}$,    which      defines      a        map
$S^{3}_{Eucl}\rightarrow S^{3}_{SU(2)}$, as  the  $SU(2)$  group  is
also a three sphere \cite{4}.   All  maps  $S^{3}\rightarrow  S^{3}$
fall  into  homotopy  classes labeled by an integer  degree  of  the
map. In the case under consideration the degree of the  map  is  the
Pontryagin index
\be
\nu  =  {1\over   16\pi^{2}}   \int   trF_{\mu\nu}\tilde{F}^{\mu\nu}
\sqrt{g}d^{4}x,                                            \label{7}
\ee
where    the    dual    tensor    is     $\tilde{F}^{\mu\nu}={1\over
2\sqrt{g}}     \varepsilon^{\mu\nu\alpha\beta}F_{\alpha\beta}\     \
(\varepsilon^{0123}=1)$. It is worth noting that  this  equation  in
fact does  not  involve  the metric, so  we  infer  that  instantons
interpolating  between  two  vacua  in  EYM  theory   may  be  still
described  in  terms  of  gauge invariant,  topological  indices  of
the YM field.

Let us introduce the Chern-Simons current
$$
K^{\mu} = {1\over  8\pi^2}  tr  {\varepsilon^{\mu\nu\alpha\beta}  \over
\sqrt{g}}  A_{\nu}(\nabla_{\alpha}A_{\beta}   -    {2{\it    i}\over
3}
A_{\alpha}A_{\beta}), $$
whose divergence is
\be
\nabla_{\alpha}
K^{\alpha}=
{1\over     16\pi^{2}}     tr     F_{\mu\nu}\tilde{F}^{\mu\nu},
                                                    \label{8}  \ee
where $\nabla_{\alpha}$ is the covariant derivative with respect  to
the spacetime metric. Pass to the $A_{0}=0$ gauge,  which  violates
the   equivalence   between   the   Euclidean   coordinates.   Using
Eqs.(\ref{7}),(\ref{8}), one obtains
\be
\nu = \left.\int
\sqrt{g}K^{0}d^{3}x\right|^{\tau=+\infty}_{\tau=-\infty}
+ \int dt\oint\vec{K}d\vec{\Sigma} =
 \left.\int
\sqrt{g}K^{0}d^{3}x\right|^{\tau=+\infty}_{\tau=-\infty}      \equiv
k(\infty)-k(-\infty).                                    \label{9}
\ee
The surface integral in this expression can be  omitted  as  the  YM
field tends to a pure gauge at the spatial infinity. Indeed,  for  a
pure gauge field taken in the $A_{0}=0$ gauge, $K^{0}$ is  the  only
non-zero component of the Chern-Simons current. So, the surface term
in Eq.(\ref{9}) vanishes when the boundary moves to infinity. In the
limit $\tau\rightarrow\pm\infty$ the gauge field  tends  to  a  pure
gauge, ${\it i}U\partial_{j}U^{-1}$, which allows one to write
\be
k = \int K^{0}d^{3}x  =  {1\over  24\pi^{2}}  \int
\varepsilon^{ijk}trU\partial_{i}U^{-1}U\partial_{j}U^{-1}U\partial_{k}
U^{-1} d^{3}x.                                            \label{10}
\ee
Thus, one  can  attach  to  any  EYM  vacuum  a  number,  $  k$   --
the Chern-Simons charge -- and the difference of two  such   numbers
for  vacua connected via an instanton is always an integer and gauge
invariant  as well.  The  value  of  $k$  itself  is  neither  gauge
invariant nor an integer, however  if  the   function   $U(\vec{x})$
in  Eq.(\ref{6})   satisfies   the  following  additional  condition
\cite{1}
\be
\lim_{|\vec{x}|\rightarrow\infty}U(\vec{x}) = 1,          \label{11}
\ee
then $k$ will be an integer and may be treated  as  the  topological
winding number. Indeed, the  condition   (\ref{11})   implies   that
$U(\vec{x})$  may  be viewed as a  map   $S^{3}_{spatial}\rightarrow
S^{3}_{SU(2)}$,   where  $S^{3}_{spatial}$   is   the   compactified
3-dimensional  space \cite{1,18}, so an integer degree  of  the  map
again appears, which is exactly given by  Eq.(\ref{10}).  Also,  $k$
will be  invariant  with  respect  to  small  gauge  transformations
generated  by  functions  obeying   (\ref{11})   and   having   zero
Chern-Simons index (\ref{10}).

Notice, that the condition (\ref{11}) arises naturally. Indeed,  for
the trivial vacuum one has $0=A_{j}=\ii  U\partial_{j}U^{-1}$, which
implies that $U(\vec{x})=1$ (up to a global  gauge  transformation),
so the condition (\ref{11}) holds. During the time evolution one has
in the $A_{0}=0$ gauge
\be
0=A_{0}(t,\vec{x})=
\lim_{|\vec{x}|\rightarrow\infty}
A_{0}(t,\vec{x})=
\lim_{|\vec{x}|\rightarrow\infty}\ \ii U\partial_{t}U^{-1},
                                                       \label{11:1}
\ee
which gives \cite{18}
\be
\lim_{|\vec{x}|\rightarrow\infty}U(t,\vec{x}) = 1.     \label{11:2}
\ee
Thus, the condition (\ref{11}) holds for any  vacuum  which  may  be
connected with the trivial vacuum via an instanton.

So, one can see that the vacuum classification  in  the  EYM  theory
resembles that arising in the flat space YM theory.  The  EYM  vacua
can be classified in terms of the integer  winding  numbers  of  the
gauge field. Gravitation does not introduce any major  peculiarities
as long as one considers only manifolds with $R^{4}$-topology.

It is natural to ask how large the amplitude of  transition  between
two EYM vacua with distinct winding numbers is. As is known, in flat
spacetime  this  amplitude  is  small  because   the   corresponding
instanton action is large \cite{1}. It is  easy  to  see   that  the
inclusion  of  gravity does not improve the situation.  Indeed,  the
inequality $$  tr(F_{\mu\nu}\pm\tilde{F}_{\mu\nu})^{2}   =   2    tr
(F_{\mu\nu} F^{\mu\nu} \pm F_{\mu\nu}\tilde{F}^{\mu\nu}) \geq  0  $$
implies that the total EYM action satisfies the following relation
\be
S^{Eucl}_{EYM}  =
{1\over   2g^{2}}\int   trF_{\mu\nu}F^{\mu\nu}\sqrt{g}
d^{4}x +  S_{G}^{Eucl}  \geq  {8\pi^{2}\over  g^{2}}|\nu|  +
S_{G}^{Eucl}.                                            \label{12}
 \ee
As the Euclidean gravitational action is positive on  shell,  it  is
clear  that the inclusion  of   gravity   may   only   lead   to  an
additional suppression  of   the   instanton   transitions   between
sectors  with distinct winding numbers.

However, as was first argued within  the  context  of  the  standard
model \cite{7}, instanton  estimates for fermion  number   violating
amplitudes may break down in the classical, many-quanta limit, where
the sphaleron  estimates  are   relevant.   One   may    reformulate
these arguments in the EYM case as follows.  Let  quarks,  $q$   and
leptons, $l$, be involved. Then it follows from  Eq.(\ref{12})  that
the  fermion number violating amplitude, $<qqql>$, is always  small.
Indeed,  such an amplitude may  be obtained  from   the   generating
functional whose path integral always gives rise to the small factor
$exp(-S^{Eucl})$.  However,    the    {\it    inclusive}  amplitude,
$<qqqlA^{n}h^{m}>$, may not be small, where  $A$ and $h$  stand  for
the  gauge bosons and gravitons, and $m$ and $n$ are  large  numbers
(see Ref.\cite{7} for details).  Notice,  that   such  an  inclusive
amplitude arises naturally in a process mediated by a sphaleron,  as
the latter, being a  classical  object,  involves  large  number  of
quanta. In  other words, one may expect that the collision  of   two
high  energy  fermions produces the sphaleron  as  an   intermediate
state.  Then  it  decays  giving  a  large  number  of   gauge   and
gravitational quanta and also other fermions as a side effect due to
the anomaly. A straightforward evaluation of the amplitude for  such
a process can be  done  with  the  use  of  perturbation  techniques
\cite{7}, but we shall proceed along  a  different  line  and  apply
instead non-perturbative approximation.


\section{An example of the  real  time  winding  number  changing
history}

Now, we return to the Lorenzian sector and  introduce  a  family  of
paths lying in the EYM configuration function space  and  connecting
homotopically distinct vacua. In the next section this will allow us
to apply Bitar and  Chang's  technique  in  order  to  represent  an
amplitude  of  the   winding   number   changing   transition.   The
considerations  in  this  section  follow  the  line   sketched   in
Ref.\cite{10}.

We shall consider spherically symmetric fields and so  we  need  the
corresponding expressions for the field potentials. Let us  use  the
following   parameterization   of    the    spherically    symmetric
gravitational field
\be
ds^{2} = R^{2}_{g} \{(1-\frac{2m}{r})\sigma^{2}dt^{2} -
\frac{dr^{2}}{1-2m/r} - r^{2}(d\vartheta^{2} + sin^{2}\vartheta
d\varphi^{2})\},                                      \label{13}
\ee
where $R_{g}=\sqrt{4\pi}{\it l}_{pl}/g$ is the natural length  scale
arising in the EYM theory with ${\it l}_{pl}$ being Planck's length,
all other quantities in Eq.(\ref{13}) being dimensionless,  and  $m$
and $\sigma$ being functions of  the  radial  coordinate,  $r$,  and
time, $t$,  as  well.  Topological  triviality  of  the  space  time
manifold implies that the metric must be regular and  asymptotically
flat; this means that $m,\ \sigma$ are smooth  functions  satisfying
the following boundary conditions
\be
\sigma\rightarrow 1,\ m\rightarrow\ const\ {\rm as}\
r\rightarrow\infty;
\ \ \
m=o(r),\   \sigma=\sigma_{0}   +   o(r)\
{\rm as}\    r\rightarrow    0.
                                                         \label{14}
\ee

The spherical YM field is given by Witten's ansatz \cite{20}, which
can be represented in the following form
\be
A = W_{0}L_{1}\ dt + W_{1}L_{1}\ dr
+\{\k_{2} L_{2} - (1-\k_{1})\ L_{3}\}\ d\vartheta +
\{(1-\k_{1})\ L_{2} + \k_{2}\ L_{3}\}sin\vartheta\ d\varphi,
                                                         \label{15}
\ee
where
$$
L_{1} = sin\vartheta cos\varphi\ (\frac{\tau_{1}}{2}) +
sin\vartheta sin\varphi\ (\frac{\tau_{2}}{2}) +
cos\vartheta\ (\frac{\tau_{3}}{2}),\ \ $$ \be
L_{2}=\partial_{\vartheta}L_{1},\ \ \ L_{3}=\frac{1}{sin\vartheta}
\partial_{\varphi}L_{1},                                 \label{16}
\ee
$W_{0},W_{1},\k_{1},\k_{2}$ are  functions  of  $r$  and
$t$.
It  is  convenient  to  use  also  another
parameterization of this ansatz introducing functions $\Omega_{0},
\Omega_{1}, f, \alpha$ as follows:
\be
W_{0}=\Omega_{0}+\dot{\alpha},\ W_{1}=\Omega_{1}+\alpha',\
\k_{1}=fcos\alpha,\ \k_{2}=fsin\alpha,    \label{omega}
\ee
where dot and  prime  denote  differentiation  with  respect  to
$t$  and  $r$ correspondingly.
One may see that only three of these  function  are  essential,  the
fourth one, $\alpha$, is actually a pure  gauge  parameter, we shall
call it an angle parameter of the field. The gauge transformation
\be
A_{\mu}\    \rightarrow\     U(A_{\mu}+     \ii\partial_{\mu})U^{-1}
                                                           \label{17}
\ee
with
\be
U = exp\{\ii\beta(t,r)L_{1}\}                          \label{18}
\ee
preserves  the  form  of  the  field  (\ref{15}),(\ref{omega})
leaving  functions
$\Omega_{0},  \Omega_{1},  f$  invariant  and  altering  the   angle
parameter as follows
\be
\alpha\  \rightarrow\  \alpha  +  \beta   .            \label{19}
\ee
Also, gauge invariant components of the  energy-momentum  tensor  of
the field (\ref{15}) (see Eq.(\ref{A2})) do not include $\alpha$.

If  one  puts  in  Eq.(\ref{15}),(\ref{omega})
$\Omega_{0}=\Omega_{1}\equiv   0,\
f\equiv 1$, allowing for the $\alpha$  to  be  arbitrary,  then  the
energy of  the  field vanishes, so  the  field  will  correspond  to
a  pure  gauge,  $\ii U\partial_{\mu}U^{-1}$, where $U$ is given  by
Eq.(\ref{18})  with  $\beta$ replaced by $\alpha$.  The  topological
vacua arise if one imposes the following additional restrictions  on
the angle parameter: $\dot{\alpha}=0$, which insures  the  $A_{0}=0$
gauge condition, and also
\be
\alpha(0)=0,\ \  \alpha(\infty)=-2\pi  k.                \label{20}
\ee
Indeed,  the  Chern-Simons  charge  for  the  field  (\ref{15})  (see
Appendix B) is
\be
k=\int\sqrt{-g}K^{0}d^{3}x = \frac{1}{2\pi}
\int_{0}^{\infty}dr
\{W_{1}(\k_{1}^{2}+\k_{2}^{2}-1)+\k_{2}\k_{1}'+
(1-\k_{1})\k_{2}'\};   \label{21}
\ee
in vacuum this expression takes the form
$$
\left. \frac{1}{2\pi}(sin\alpha-\alpha)\right|_{0}^{\infty},
$$
which  is  the  integer,   $k$,   provided   that   the    condition
(\ref{20})  holds  (the  condition  $\alpha(0)=0$  in  Eq.(\ref{20})
insures regularity  of  the field at the origin). It is  clear  then
that  the  function   (\ref{18})   with   time-independent   $\beta$
vanishing both at the origin  and  infinity  generates  small  gauge
transformations as the condition (\ref{20})  remains unchanged.

Summarizing,    we     conclude     that     the     gauge     field
(\ref{15}),(\ref{omega})  with  $\Omega_{0}=  \Omega_{1}  =f-1\equiv
0$  and  arbitrary  time-independent  $\alpha$ obeying the condition
(\ref{20}),    represents    the    topological     vacuum,     $\ii
U\partial_{j}U^{-1}$ with $U=exp(\ii\alpha L_{1})$, and the  winding
number is $k$.

We consider  now an explicit   example   of   the   winding   number
changing evolution of the gauge field \cite{10}:
\be
A_{\mu}(t,\vec{x})=\frac{\ii}{2}(1-K(r))U\partial_{\mu}U^{-1},
\ \ \ \
U=exp(2\ii\lambda L_{1}),                                \label{22}
\ee
where  $\lambda$  is  some  function  of  time,   $\lambda(t_{0})=0,
\ \lambda(t_{1})=\pi$, where $t_{0}<t_{1}$, and $K(r)$ satisfies
\be
K(0)=1,\ \ K(\infty)=-1.                                  \label{23}
\ee
It is clear that the field  (\ref{22})  is   spherical   and   hence
it  can  be represented in the form  (\ref{15}) giving for the
corresponding functions
\be
W_{0}=(1-K)\dot{\lambda},\ W_{1}=0,\
\k_{1} =cos^{2}\lambda+Ksin^{2}\lambda,\
\k_{2} =(1-K)sin\lambda cos\lambda,                     \label{23:1}
\ee
and also
$$
\Omega_{0}=-\dot{\lambda}\frac{K(K^{2}-1)}{K^{2}+ctg^{2}\lambda},
\ \ \Omega_{1}=\frac{K'ctg\lambda}{K^{2}+ctg^{2}\lambda}, $$
\be
f^{2}=sin^{2}\lambda K^{2}+cos^{2}\lambda,\ \ \
tg\alpha=\frac{(1-K)ctg\lambda}{K+ctg^{2}\lambda}.     \label{24}
\ee
One may see that the field (\ref{22})  vanishes  in  the  beginning,
$\lambda=0$, as well as at the end, $\lambda=\pi$, of the  evolution
(provided that $\dot{\lambda}(t_{0})=\dot{\lambda}(t_{1})=0$). Thus,
at  first  sight,  this  field  corresponds  to  a   loop   in   the
configuration function space interpolating between the same  trivial
vacuum. However, the crucial issue here is the right choice  of  the
gauge.  The  homotopical  vacuum  classification  is  based  on  the
$A_{0}=0$ gauge condition  \cite{1},\cite{18},\cite{jackiw},  so  one
has to  pass  to  this  gauge.  To  do  this,  perform   the   gauge
transformation (\ref{19}) and demand
\be
A_{0}=(\Omega_{0}+\dot{\alpha}+\dot{\beta})L_{1}   =0,    \label{25}
\ee
where $\Omega_{0}$ and $\alpha$ are given by Eq.(\ref{24}). In  this
way  we  obtain  the  equation  for  the  gauge  parameter  $\beta$:
\be
\Omega_{0}+\dot{\alpha}+\dot{\beta}=\dot{\lambda}(1-K)+
\dot{\beta}=0,                                     \label{25:1}
\ee
which yields
\be \beta=(K-1)\lambda + B(r),                     \label{25:2}
\ee
where the function $B(r)$ may  be  put  equal  to  zero  using  the
residual freedom  of  the  time-independent  gauge  transformations,
which preserve the $A_{0}=0$  gauge  condition.  Thus,  in  the  new
gauge,  the  invariant  functions  $\Omega_{0},  \Omega_{1},f$   are
still determined by Eq.(\ref{24}), but the angle parameter  $\alpha$
is replaced by
\be
\alpha_{0}= (K-1)\lambda +
arctg\frac{(1-K)ctg\lambda}{K+ctg^{2}\lambda}.           \label{26}
\ee
The corresponding expressions for  the  functions  $W_{0},W_{1},
\k_{1},\k_{2}$ in the new gauge are
$$
W_{0}=0,\ W_{1}=\lambda K',\
\k_{1}=cos\lambda cosK\lambda+Ksin\lambda sinK\lambda,\ $$\be
\k_{2}=cos\lambda sinK\lambda-Ksin\lambda cosK\lambda.    \label{27}
\ee
One can see that the field (\ref{15}) specified by  these  functions
vanishes at $\lambda=0$. On the  other  hand,  at  the  end  of  the
evolution, at $\lambda=\pi$, the field (\ref{15}),(\ref{27}) is  the
non-trivial pure gauge. The angle parameter of the new vacuum can be
defined from Eq.(\ref{26}):
\be
\alpha_{0} (r) = \pi(K(r)-1).                        \label{28}
\ee
One may see also from  Eqs.(\ref{20}),(\ref{23})  that  this  indeed
corresponds  to a vacuum with unit winding number.

Thus the  field  (\ref{22})  indeed  interpolates  between  distinct
vacua. To show a change of the winding number during  the  evolution
(\ref{22}) one has to estimate the Chern-Simons number as a function
of $\lambda$ (see also \cite{brodbeck} for discussion of the subject
in the case when arbitrary gauge  groups  are  involved).  Inserting
functions (\ref{27}) into Eq.(\ref{21}) and using Eq.(\ref{23}),
one obtains
$$
k=\frac{1}{2\pi}\int_{1}^{-1}dK
(-\lambda+sin\lambda cos\lambda+
\lambda cos\lambda cosK\lambda- sin\lambda cosK\lambda +$$\be
+K\lambda sin\lambda sinK\lambda)
=\frac{1}{\pi}(\lambda - sin\lambda cos\lambda ),    \label{29}
\ee
which varies from zero to unity as  $\lambda $  goes  from  zero  to
$\pi  $.  It  is  worth   noting   that the   middle   of    the
evolution,  $\lambda  =\pi  /2$,  corresponds  to   a   half-integer
Chern-Simons  number, $k=1/2$ \cite{10}. It is clear  now  that  the
boundary conditions (\ref{23}) for the function $K$ in  Eq.(\ref{22})
are important. Imposing instead the conditions
\be K(0)=K(\infty)=1,                                \label{even}
\ee
the value of the integral in the right hand  side  of  Eq.(\ref{29})
would be zero as would be the Chern-Simons number. So, the  boundary
conditions (\ref{23}) insure that the field (\ref{22})  interpolates
between homotopically different vacuum sectors. On the  other  hand,
the conditions (\ref{even}) would make the field interpolate between
vacua inside the same vacuum equivalency class.

Now, being aware  that the field  (\ref{22})   indeed   relates   to
the  transition  between  distinct  vacua,  we  adopt   a   slightly
different  interpretation of this example. Let us treat the quantity
$\lambda$ in Eq.(\ref{22})  not as a function  of  time  but  rather
just as a parameter running from  zero  to $\pi$.  This  means  that
the field (\ref{22}) is viewed now as a   one-parameter  family   of
static   field   configurations   or   a    path    in    the    EYM
configuration function space.

Next, we want to specify a gravitational field related to this path.
The most natural choice is for the  matter  field (\ref{22}) and the
corresponding gravitational field to form an admissible set  of  the
initial data on an initial time-symmetry  hypersurface. So we impose
the initial value constraints, from which the only  non-trivial  one
in our case  is Einstein's  equation  (\ref{A9}).   For   the   sake
of convenience only, we impose also the equation (\ref{A11}),  which
specifies the  lapse  function  $\sigma$.  Using   Eqs.  (\ref{A2}),
(\ref{24}), (\ref{A9}), (\ref{A11}), we get the following  equations
for the spherical gravitational field related to the path (\ref{22})
$$
m'=sin^{2}\lambda\{(1-\frac{2m}{r})K'^{2}      +      sin^{2}\lambda
\frac{(K^{2}-1)^{2}}{2r^{2}}\}, $$
\be
\sigma'=2sin^{2}\lambda\frac{K'^{2}}{r}\sigma.           \label{32}
\ee
Straightforward   integration   of   these    equation    with   the
boundary conditions  (\ref{14})  allows  us  to  express   $m$   and
$\sigma$  in  terms  of the quantities, which parameterize the field
(\ref{22}) \cite{10}:
$$
\sigma  (r)   =   exp\{   -2sin^{2}\lambda\int_{r}^{\infty}   K'^{2}
\frac{dr}{r}\}, $$
\be
m(r)=\frac{sin^{2}\lambda}{\sigma(r)}\int_{0}^{r}     (K'^{2}      +
sin^{2}\lambda\frac{(K^{2}-1)^{2}}{2r^{2}})\sigma dr.    \label{33}
 \ee
If  $K(r)$  is  a  sufficiently  smooth    function    then    these
expressions specify a  metric,  which  is  asymptotically  flat  and
regular   for   any  $\lambda$   and   takes   the   vacuum   value,
$\eta_{\mu\nu}$, at the  ends  of  the  path,  $\lambda=0,\pi$.  The
corresponding (dimensionless) ADM mass is
\be
M = \lim_{r\rightarrow\infty}m(r).                       \label{34}
\ee
Recall now that $K(r)$ in the above expressions  may  be  arbitrary.
This means that Eqs.(\ref{22}),(\ref{33}) define not  one  path  but
rather a  family of vacuum-to-vacuum paths
\be
 \{A_{j}(\vec{x},     \lambda);    \left.        g_{\mu\nu}(\vec{x},
\lambda)\}
\right|_{K(r)}, \ \ \ \lambda\in [0,\pi]                 \label{35}
\ee
in the EYM configuration function space, each  path  of  the  family
being  specified  by  a   function   $K(r)$   satisfying  conditions
(\ref{23}).  The energy (\ref{34}) for each path  vanishes  for  the
vacuum values, $\lambda=0,\pi $, and has a  maximum  in  between  at
$\lambda =\pi /2 $.

Now, one can see that the fields (\ref{22}),(\ref{33}) given at  the
point  $\lambda=\pi/2$  coincide  with  the  solutions  of  the  EYM
equations found by Bartnik and McKinnon (BK) \cite{9} provided  that
the  function  $K(r)$   is  identified  with  the  corresponding  BK
magnetic amplitude $w_{n}(r)$ (see Appendix B). If the number $n$ is
odd, then $K(r)=w_{n}(r)$ obeys the boundary conditions  (\ref{23}),
so odd-$n$ BK solutions belong to paths  connecting  distinct  vacua
(even-$n$  solutions   belong   to   paths   interpolating   between
representatives of the same vacuum class  as  $w_{n}$  functions  in
this case obey the condition (\ref{even})).  This  circumstance  has
allowed us initially  to  interpret  the  odd-$n$  BK  solutions  in
Ref.\cite{10} as sphalerons, because they relate to  the  top of the
potential  barrier  dividing  homotopically  distinct   EYM   vacua,
$\lambda =\pi/2$,  and  also have the Chern-Simons  index  one-half.
However, as is mentioned in  the Introduction (see also  \cite{10}),
the   analogy   with   the   standard   YMH  sphaleron  \cite{5}  is
incomplete because none of the BK solutions relate to  an   absolute
minimum barrier  height,  and  hence  an additional investigation is
required  in  order  to  establish  the sphaleron  nature  of  these
solutions.

\section{The  winding   number   changing   transition
amplitude.}

In this section we represent the amplitude  of a quantum  transition
between homotopically non-equivalent EYM vacuum sectors at  non-zero
energy {\it a la} Bitar and Chang. We  assume  that  the  transition
occurs only along the  paths  (\ref{22}),(\ref{33}),  which  is  our
basic approximation. Notice that Bitar and Chang's approach gives an
adequate description of the transition process provided one is  able
to take into account {\it all} VTV paths.  But  the   main advantage
of  the  method,  perhaps,    is   that   it   may   give   a   good
description of the transition even if one takes only some particular
set of the vacuum-to-vacuum paths \cite{11}.

To obtain the transition amplitude, let us first choose a path  from
the family (\ref{22}),(\ref{33}).  To  find  the  partial  amplitude
related to this path, allow for the parameter $\lambda$ to depend on
time: $\lambda\rightarrow\lambda(t)$, and calculate  the  action  of
the  fields.  Inserting  Eqs.(\ref{22}),(\ref{33})  with   $\lambda=
\lambda(t)$ into Eq.(1), represent the action in the form
\be
S_{EYM}=\frac{4\pi}{g^{2}}\int (L_{G}+L_{YM})\ dt,      \label{38}
\ee
where the   Lagrangians
$(4\pi/g^{2})L_{G}$ and $(4\pi/g^{2})L_{YM}$ of
gravitational   and   the   YM   field are  the spatial
integrals
of    the    corresponding    Lagrangian    densities    given    by
Eqs.(\ref{2}),(\ref{3}). Consider first the  matter  part   of   the
action.  Using Eqs.(\ref{3}), (\ref{24}), (\ref{A3}) and taking
 into
account  the  Einstein equations (\ref{32}), one obtains
\be
L_{YM}=\frac{\mu(\lambda)}{2}\dot{\lambda}^{2}-\iii m'\sigma dr,
                                                        \label{39}
\ee
where
\be
\mu(\lambda)=\iii\frac{r^{2}}{\sigma}(K'^{2}+2sin^{2}\lambda\
\frac{(K^{2}-1)^{2}}{r^{2}-2mr\ })dr,                    \label{40}
\ee
quantities  $m,\  m',\  \sigma$  being  given   by   Eqs.(\ref{32}),
(\ref{33})  and $K(r)$ being a function specifying  the  path  under
consideration.

The calculation of the gravitational part of the action is performed
in the Appendix A. For regular geometry described  by  Eq.(\ref{13})
the horizon term in Eq.(\ref{fin}) disappears, so one gets
\be
L_{G}=-\int_{0}^{\infty} m\sigma'dr                    \label{42}
\ee
(when  one  passes  in  Eq.(\ref{fin})  from  dimensional  variables
$t,r,m$ to dimensionless ones, which  are  used  throughout  in  the
paper,  the   multiplier   $4\pi/g^{2}$   explicitly   depicted   in
Eq.(\ref{38}) arises).

One can see that the integral entering Eq.(\ref{42})  combines  with
that coming from Eq.(\ref{39})   to   give   the   total  derivative
under integration. This allows us to represent the total EYM  action
in the following form
\be
S_{EYM}=\frac{4\pi}{g^{2}}\int
(\frac{\mu(\lambda)}{2}\dot{\lambda}^{2}-U(\lambda))\ dt,\label{43}
\ee
where  the  effective  mass  term,  $\mu(\lambda)$,  is   given   by
Eq.(\ref{40}),
and the potential, $U(\lambda)$, reads
\be
U(\lambda)=\iii (m\sigma)'dr = m(\infty).               \label{44}
\ee
This coincides with the ADM mass given by   Eq.(\ref{34})  (that  is
true provided that the geometry is everywhere regular; see  Appendix
A). Explicitly one has
$$
U(\lambda)\equiv U[K(r),\lambda]=$$
\be
=sin^{2}\lambda\iii (K'^{2} +
sin^{2}\lambda\frac{(K^{2}-1)^{2}}{2r^{2}})
exp(-2sin^{2}\lambda\int_{r}^{\infty}K'^{2}\frac{dr}{r})dr.
                                                         \label{45}
\ee

One may see that the result obtained corresponds   to   the   action
of   an    effective    particle    with  position-dependent   mass,
$\mu(\lambda)$, moving  in  a  one-dimensional  external   potential
$U(\lambda)$.  The  potential has the  typical  barrier  shape:  for
each $K(r)$ it  vanishes for the vacuum values, $\lambda=0,\pi$, and
reaches  a  maximum  in between at $\lambda=\pi/2$ (the  latter  can
be seen if we pass to a new  independent  variable  $z=r/sin\lambda$
under the integration in Eq.(\ref{45})).

Thus, we arrive at the one-dimensional barrier  transition  problem.
The corresponding one-dimensional Schr\"odinger equation then reads
\be
{\cal H}=\frac{p^{2}}{2\mu(\lambda)}+U(\lambda)=E,
                                                        \label{37}
\ee
with $p$ being the momentum conjugated to $\lambda$,  and  the
quantity $E$ has the sense of the energy of the asymptotically  free
quantum states (the operator ordering problem can be avoided in this
case \cite{11}). This allows us  to  write  down  the  corresponding
partial WKB transition amplitude as follows
\be
{\cal A}_{K(r)}=Bexp\{ \ii\frac{4\pi}{g^{2}}\Phi[K(r)]\}=
Bexp\{ \ii\frac{4\pi}{g^{2}}
\int_{0}^{\pi}  d\lambda \sqrt{2\mu(\lambda)[E-U(\lambda)]}\},
                                                          \label{46}
\ee
where $B$ absorbs all other WKB factors, which  are  inessential for
our present considerations. The subscript $K(r)$ indicates that this
partial amplitude  relates   to   a   single   path   specified   by
$K(r)$. If the quantity $E-U(\lambda)$ is negative, then one  should
take the value of the square root lying in the  upper  half  of  the
complex plane.

The last step in finding the transition amplitude  is  to  represent
the  total  amplitude  as  the  sum  over  all  partial   amplitudes
(\ref{46}):
\be
{\cal A}=
\sum_{K(r)}{\cal A}_{K(r)}.                               \label{47}
\ee
This expression gives the amplitude of the winding  number  changing
transition in the EYM theory at arbitrary energy $E$. It  is  clear,
however, that taking the sum is virtually impossible. The  only  way
to estimate this  sum  is  to  make  use  of  the  stationary  phase
approximation. However, as was shown in Ref.\cite{11},  the  finding
of the exact stationary phase path implies solving a system  of  the
coupled differential equations, which,  unfortunately,  lies  beyond
our abilities.

To proceed further  we  will  not   calculate   the  amplitude,  but
consider instead  a  more  simple  problem. Namely, we want to  find
out under which conditions this amplitude will not be small.

We use the following terminology. Let the  energy,  $E$,  be  fixed.
Consider a path (\ref{22}),(\ref{33})  that  lies  entirely  in  the
classically allowed region:
\be
U[K(r),\lambda]<E,\ \  \lambda\in [0,\pi].               \label{48}
\ee
Such a path will be called {\it an  overbarrier  path}.  If  a  path
passes also through the classically forbidden region then  we  shall
call it an {\it underbarrier path}, for the evolution  along  such a
path  implies barrier penetration.

For an underbarrier paths the amplitude (\ref{46}) is small,  as  it
includes the small tunneling factor. This allows us to exclude  from
the sum (\ref{47}) the contribution of all  underbarrier  paths,  as
this contribution  is  certainly small. The remaining  sum  over the
overbarrier paths will not be small if only this sum includes a term
(or terms),  ${\cal  A}_{K(r)}$,  whose  value  is  stationary  with
respect to small variations  of  $K(r)$. Such a term corresponds  to
the contribution of a stationary  phase  path. Thus,  the  amplitude
(\ref{47}) will not be small provided  there  exists  a   stationary
phase path on the set  of  all  overbarrier  paths.  Therefore,   to
proceed further, we first need  to  select  the  overbarrier   paths
from  the  set  (\ref{22}),(\ref{33}),  and  next  to  look  for   a
stationary phase path among them.

\section{Structure of the energy surface}

To select the overbarrier paths in accordance with Eq.(\ref{48}) one
has to know properties of the  potential  $U[K(r),\lambda]$.  It  is
useful to treat the functional   $U[K(r),\lambda]$   in  geometrical
terms,  viewing  it  as  an  infinite  dimensional  surface  in  the
corresponding function space. Let us  call  this   surface  {\it  an
energy  surface}  or  {\it  a  potential  barrier  surface}.  It  is
natural to call the directions  in  this  surface    generated    by
$\partial/\partial\lambda$    and   $\partial/ \partial K(r)$   {\it
a  transverse  direction}  and  {\it  a   longitudinal   direction}
respectively.  Indeed,  changing  of  $\lambda$  with  fixed  $K(r)$
corresponds to motion across  the  barrier  towards  a  neighbouring
vacuum. Changing of $K(r)$ implies passing  to  other   paths,  i.e.
the motion along  the   barrier.   When   $K(r)$   is   fixed,   the
potential    reaches     a     maximum      at      $\lambda=\pi/2$,
so  we  will  call the functional
\be \varepsilon[K(r)]=U[K(r),\lambda=\pi/2]                \label{49}
\ee
{\it a barrier height  functional}  and  say  that  it  defines  the
profile of the  top  of  the barrier.

Consider the critical  points of the energy  surface.  It  is  clear
that such points have to belong to the top, $\lambda=\pi/2$.  Direct
variation of  the  functional  (\ref{49})  yields  (see Appendix C)
\be
\delta\varepsilon=2\iii\{-((1-\frac{2m}{r})\sigma
K')'+\frac{K(K^{2}-1)} {r^{2}}\sigma\}\delta K  dr,      \label{50}
\ee
where $\delta K(0)=\delta  K(\infty)=0$,  functions  $m,\sigma$  are
given by Eq.(\ref{B3}). The vanishing  of  this  variation  requires
the condition
\be
((1-\frac{2m}{r})\sigma K')'=\sigma\frac{K(K^{2}-1)}{r^{2}}
                                                        \label{51}
\ee
to hold, which is exactly the  YM   equation   (\ref{A12}).   Noting
then, that  the  functions  $m,\sigma$  in  this  case  satisfy  the
equations
\be
\sigma'=2\frac{K'^{2}}{r}\sigma,\ \ \ \ \ \ \ \ \ \
 (m\sigma)'=(K'^{2}+\frac{(K^{2}-1)^{2}}{2r^{2}})\sigma, \label{Bb}
\ee
which    are     identical     to     the     Einstein     equations
(\ref{A13}),(\ref{A14}),    we    conclude   that   the    condition
$\delta\varepsilon=0$  insures  the  on-shell   condition    $\delta
S_{EYM}=0$. All points of the energy surface relate to  the  regular
field configurations, hence, critical points of the top  relate   to
regular solutions of the EYM equations, i.e. to the EYM  sphalerons.
One  may say that the EYM sphalerons  ``lie''  on  the  top  of  the
potential  barrier. (Of course, these arguments only  establish  the
coincidence of the  extrema of the action and those of the truncated
mass functional under  consideration.  When  all  field  degrees  of
freedom are involved rather than only those  which  are  spherically
symmetric, the correspondence between extrema of the energy  and the
action is not proven, although it is  very  likely  that  it  indeed
holds; see Ref.\cite{21}.)

Next, we consider the local properties  of  the  energy  surface  in
the vicinity of a critical  point.  Firstly,  we   know   that   the
potential     energy     decreases     when     the      transversal
$\lambda$-coordinate deviates from  the  $\lambda=\pi/2$  value,  so
locally, near any critical point, the energy surface  is  a  saddle,
possessing at least one  negative  direction.  Therefore,  each  EYM
sphaleron has at least one negative mode which we will refer  to  as
{\it transverse rolling down mode} \cite{22}. In the YMH theory this
 mode  is  the single negative  mode   of
the  sphaleron   (provided that the   Higgs field  self-coupling  is
not  too  large  \cite{13}),  so  the  energy  increases   in    any
longitudinal  direction. In the EYM theory  there  exist  additional
sphaleron negative modes, for the potential energy decreases also in
some longitudinal  directions.  Indeed,  direct   calculation   (see
Appendix C) shows that in the vicinity of a sphaleron the  following
expansion holds
\be
\varepsilon [w_{n}(r)+\varphi(r)]=
\varepsilon[w_{n}(r)]+\delta^{2}\varepsilon +\ldots ,
                                                       \label{52}
\ee
where the first order term vanishes and dots denote higher order
terms. The second order term reads
\be
\delta^{2}\varepsilon =  \iii\f(-\frac{d^{2}}{dr_{\ast}^{2}}  +V)\f\
dr_{\ast}, \label{53}
\ee
with the new radial coordinate, $r_{\ast}$,  being  defined  by  the
following   relations    $\frac{dr}{dr_{\ast}}=\sigma(1-2m/r)$,    $
r_{\ast}(r=0)=0$,    the    perturbation    $\f(r_{\ast})$     obeys
$\f(0)=\f(\infty)=0$. The effective potential is
\be
V=\sigma^{2}(1-\frac{2m}{r})\{\frac{3w_{n}^{2}-1}{r^{2}} +
\frac{8}{r^{3}}w_{n}'w_{n}(w_{n}^{2}-1)-
\frac{4}{r^{2}}w_{n}'^{2}(1-\frac{(w_{n}^{2}-1)^{2}}{r^{2}})\},
                                                         \label{54}
\ee
where  functions  $m$  and  $\sigma$  relate  to  the  corresponding
sphaleron solutions, the derivatives are calculated with respect  to
$r$. One may see that the scalar product for different  perturbation
modes        can        be        naturally        defined        as
$<\f_{1},\f_{2}>=\int\f_{1}\f_{2}dr_{\ast}$; independent modes being
orthogonal. It is obvious  that  if  the  differential  operator  in
Eq.(\ref{53}) has negative eigenvalues,
\be
(-\frac{d^{2}}{dr_{\ast}^{2}}+V)\     \f\     =\omega^{2}\f,\      \
\omega^{2}<0,                                         \label{55}
\ee
corresponding  eigenfunctions,   $\f$,    will    specify  the  {\it
longitudinal} negative directions on the energy  surface,  that  is,
the longitudinal negative sphaleron  modes.  It  is   worth   noting
that Eq.(\ref{55}) coincides (up to a  constant   multiplier)   with
that  first obtained by Straumann and  Zhou   \cite{23}   in   their
analysis  of the linear stability of the BK solutions, provided that
one identifies  the  quantity $\omega$  with  the time frequency  of
perturbations   of   the  background   BK  solution.  (Notice,  that
Straumann  and  Zhou  derived  Eq.(\ref{55})  linearizing  the   EYM
equation near a background equilibrium solution. It is  interesting,
that  one  may  obtain   the   same   equation   from   the   energy
considerations. A similar  situation  is  also  encountered  in  the
theory of relativistic stars \cite{harrison},  where  the  dynamical
pulsation equation for a star composed from  perfect  fluid  can  be
derived by making use of the two different approaches.)

It is known \cite{24}, that the equation (\ref{55}) has exactly  $n$
independent negative eigenvalue solutions. Thus in the  vicinity  of
the  $n$-th  sphaleron  the  energy  surface  has,  apart  from  one
transversal negative mode, also $n$ longitudinal negative modes. The
total  number  of the negative  modes,  $n+1$, is  in agreement with
the result by Sudarsky and Wald \cite{21}.

It is obvious now that when  the  energy  $E$  exceeds  an  extremal
(non-zero) value of the barrier height, that is,  the  mass  of  the
$n$-th EYM sphaleron,
\be
E>\varepsilon[w_{n}],                                  \label{e}
\ee
then the condition (\ref{48}) is satisfied provided that  $K(r)$  is
close to $w_{n}(r)$, i.e. $K(r)=w_{n}(r)+\f(r)$ with  $\f$  being  a
small perturbation. This means that the  path  passing  through  the
$n$-th    sphaleron    (such    a     path     is     defined     by
Eqs.(\ref{22}),(\ref{33})   with   $K(r)=w_{n}(r)$)   as   well   as
neighbouring paths will be overbarrier.


\section{The stationary phase path}

Let the condition (\ref{e}) hold.  Consider  the  overbarrier  paths
defined in the preceding  section.  Now,  we  are  looking  for  the
stationary phase path on the set of these overbarrier paths. Our aim
is to show that the path passing through the sphaleron will be  such
a  stationary  phase  path.  This  means  that  the  quantum   phase
$\Phi[K(r)]$ defined by Eq.(\ref{46}) with $K(r)=w_{n}(r)$ should be
stationary on variations of $K(r)$. To check the stationarity of the
phase let us consider an arbitrary variation of $K=w_{n}$:
\be
K(r)=w_{n}(r)+\alpha\f(r),                        \label{var}
\ee
where $\alpha$ is the variational parameter,  and  the  perturbation
obeys  $\f(r)=O(r^{2})$  as   $r\rightarrow0$,   $\f(r)=O(1/r)$   as
$r\rightarrow\infty$.    Inserting     this   into     $\Phi[K(r)]$,
$U[K(r),\lambda]$     and     $\varepsilon[K(r)]$     defined     by
Eqs.(\ref{46}),(\ref{45}) and (\ref{49}) respectively,  one  obtains
the phase $\Phi_{\varphi}(\alpha)$, (index $\varphi$ refers  to  the
choice  of  the  perturbation),  and  the  one  and  two-dimensional
sections of the energy  surface:  $\varepsilon_{\varphi}  (\alpha)$,
$U_{\varphi}  (\alpha,\lambda)$.  The  phase  $\Phi[K(r)]$  will  be
stationary if $\Phi_{\varphi}(\alpha)$ has an extremum for {\it any}
$\f(r)$ in Eq.(\ref{var}).

One should note that, strictly speaking, the exact stationary  phase
path is not contained  in  the  path  family  (\ref{22}),(\ref{33}).
Finding such a path implies solving the general equations of  motion
\cite{11}, which lies beyond the  scope  of  our  present  analysis.
Paths (\ref{22}),(\ref{33})  specified  by  $K=w_{n}$  may  be  only
approximately stationary. This  means  that  for  these  approximate
paths and any independent perturbations $\varphi(r)$ in (\ref{var}),
the positions of the extremum of the corresponding  phase  functions
$\Phi_{\varphi}(\alpha)$ do not coincide exactly with  the  position
of the sphaleron, $\alpha=0$, however they are  close  together.  In
other  words,  functional  derivatives  $\left.  \delta   \Phi[K(r)]
\right|_{K=w_{n}}$ although do not vanish exactly, are  nevertheless
small.

Consider first such perturbations $\f(r)$ which increase the barrier
height:  $\varepsilon(\alpha\neq   0)>\varepsilon(0)$;   there   are
infinitely many such perturbations. One may see that  the  phase  is
stationary with respect to all these perturbations. Indeed, in  this
case  the  corresponding  two-dimensional  section  of  the   energy
surface, $U_{\varphi}(\alpha,\lambda)$, has the typical saddle shape
shown in Fig.1.  The  saddle  negative  $\lambda$-direction  on  the
picture (shown by the horizontal  arrow)  specifies  the  transverse
rolling down mode of the sphaleron \cite{22}. It is clear from  this
picture that the path passing through the sphaleron is  the  minimal
potential energy path interpolating between distinct  vacua,  so  it
corresponds to an extremum of the  phase.  The  analogous  situation
takes  place  in  the  YMH  theory  where  the  sphaleron   is   the
highest-energy point on a minimal-energy  path  connecting  distinct
vacua (at least for small values of the scalar  field  self-coupling
\cite{13}),  which implies the stationarity of the phase for such  a
path \cite{5}. Direct numerical inspection shows that,  for  several
ground state sphaleron energy-increasing perturbation modes $\f (r)$
checked,  the  phase  $\Phi_{\f}(\alpha)$  indeed  has  an  extremum
(minimum) in the vicinity of zero value of $\alpha$.

However, contrary to the situation in  the  YMH  case,  in  the  EYM
theory  not  all  perturbations  (\ref{var})  increase  the  barrier
height, for EYM sphalerons possess also longitudinal negative modes.
Inserting such perturbation modes in Eq.(\ref{var}), one obtains the
corresponding two-dimensional sections of the energy  surface  which
have the typical shape shown in Fig.2. One may see from this picture
that, for such a section, the path passing through the sphaleron  is
not the minimal energy path. So,  with  respect  to  these  negative
perturbations, the stationarity of the phase is not obvious  (it  is
usually assumed that a sphaleron  has  to  have  one  and  only  one
negative  mode  in  order  to  be  significant  for  the  transition
processes \cite{13}). Nevertheless, we are  able  to  demonstrate
the stationarity of the phase also in this case, at  least  for  the
ground state $(n=1)$ sphaleron.

For the ground  state  sphaleron  there  exists  only  one  negative
eigenmode to Eq.(\ref{55}). It  turns  out  that  the  corresponding
perturbation (\ref{var}) may be related  to  the  rescaling  of  the
sphaleron field (the scaling arguments in the EYM theory  have  also
been  considered   in   Ref.\cite{heusler}.   Namely,   instead   of
Eq.(\ref{var}) let
\be  K(r)=w_{1}(\beta r),                             \label{56}
\ee
where  $\beta$  is  the  scaling  parameter.   Inserting   this   in
Eq.({\ref{45}) one obtains the following two dimensional section  of
the energy surface:
\be
U(\beta,\lambda)=
\beta\sin^{2}\lambda\iii (w_{1}'^{2}+ \sin^{2}\lambda
\frac{(w_{1}^{2}-1)^{2}}{2r^{2}})exp(-2\beta^{2}
sin^{2}\lambda\int_{r}^{\infty}w_{1}'^{2}\frac{dr}{r})dr, \label{57}
\ee
(here $w_{1}=w_{1}(r)$), the value $U(1,\pi/2)$ being the  sphaleron
mass. This function tends to zero when  $\beta\rightarrow  0,\infty$
as well as when $\lambda\rightarrow 0,\pi$, so  one  may  see that
the potential barrier indeed can be arbitrarily low. The  plot  of
this function  is  depicted  in  Fig.2.  This  picture  demonstrates
explicitly the existence of the two independent  sphaleron  negative
modes. One such mode is the transverse rolling down  mode  \cite{22}
(the  $\lambda$-direction  on  the  picture),  the  other   is   the
longitudinal negative mode \cite{23} (the $\beta$-direction).  (Note
that     the      infinitesimal      scaling      mode,      $\left.
\f=\partial_{\beta}w_{1}(\beta r)\right|_{\beta=1}=rw'_{1}(r)$,   is
not  the  exact  eigenmode  for  the  Eq.(\ref{55}),  but  rather  a
superposition of the true negative eigenmode and some other mode).

The scaling behaviour  of  the  quantum  phase  is  defined  by  the
inserting (\ref{56}) into (\ref{46}):
\be
\Phi_{scale}(\beta)=\int_{0}^{\pi}
d\lambda \sqrt{2\mu(\beta,\lambda)[E-U(\beta,\lambda)]},
                                                     \label{phase}
\ee
where $U[\beta,\lambda]$ is given by Eq.(\ref{57}) and
\be
\mu(\beta,\lambda)=
\frac{1}{\beta}\iii \frac{r^{2}}{\sigma}
(w_{1}'^{2}+2sin^{2}\lambda\
\frac{(w_{1}^{2}-1)^{2}}{r^{2}-2\beta mr\ })dr,      \label{60}
\ee
with
$$
\sigma(r)=exp\{-2\beta^{2}sin^{2}\lambda\int_{r}^{\infty}w_{1}'^{2}
\frac{dr}{r}\}, $$
\be
m(r)=\frac{\beta\sin^{2}\lambda}{\sigma(r)}\int_{0}^{r}(w_{1}'^{2}+
sin^{2}\lambda\frac{(w_{1}^{2}-1)^{2}}{2r^{2}})\sigma dr.\label{59}
\ee
Now, one  may  see  that  the  phase  (\ref{phase})  indeed  has  an
extremum.  First  note  that  when  $\beta$  is  small,  the   phase
$\Phi_{scale}(\beta)\sim  1/\sqrt{\beta}$,  i.e.  it  diverges  when
$\beta\rightarrow 0$ (remember  that  the  energy  $E$  exceeds  the
sphaleron mass, that is, the maximal value of the potential drawn in
Fig.2). Also, and this is the crucial  point,  $\Phi_{scale}(\beta)$
diverges when $\beta$ tends to some value $\beta_{c}=1.465$, because
the integral entering Eq.(\ref{60})  diverges  in  this  limit.  The
value $\beta=\beta_{c}$ corresponds  to  such  a  rescaling  of  the
sphaleron field when the rescaled configuration begins to acquire an
event horizon (all  of  the  paths  (\ref{22}),(\ref{33}),(\ref{56})
with $\beta> \beta_{c}$ pass through virtual black holes and  give
an infinite contribution to the action). At the horizon the quantity
$r^{2}-2\beta  mr$  entering  the   denominator   in   Eq.(\ref{60})
vanishes, so the integral diverges as does the phase. So,  somewhere
in between, $0<\beta<\beta_{c}$, the phase must have a minimum  (see
Fig.3).

Thus we can see that when the energy exceeds the  ground  state  EYM
sphaleron mass, the  path  passing  through  the  sphaleron  insures
extremality  of  the  quantum  phase  with  respect  to  any   small
variations  inside  the  path  family   (\ref{22}),(\ref{33}).   The
existence of this extremum insures that the  sum  in  Eq.(\ref{47}),
being evaluated via stationary  phase  approximation,  will  be  not
small, as it will include the contribution of  a  stationary  point.
Therefore, when the energy is large,  the  winding  number  changing
transition in the EYM theory is not suppressed.

It turns out that the analysis carried  out  above  for  the  ground
state sphaleron remains valid for the higher (odd $n>1$) sphalerons.
This means that each higher sphaleron introduces a stationary  phase
path, only if the corresponding phase will be stationary  also  with
respect  to  the  additional  $n-1$  longitudinal  higher  sphaleron
negative modes, which is very plausible. Thus, if the energy exceeds
the  masses  of  all  the  EYM  sphalerons  (i.e.   $E>1$   in   the
dimensionless units used), the sum  in  Eq.(\ref{47})  includes  the
contributions of (infinitely) many stationary points (one point  for
each higher sphaleron), which may lead to an additional  enhancement
of the transition rate.

When the energy is less then the ground state  sphaleron  mass,  the
situation changes. The overbarrier paths in this case also exist, as
can be seen in Fig.2. However,  there  exists  no  stationary  phase
path, so the overbarrier passages are suppressed by the  destructive
interference. This is always occurs also in the flat space limit,
$G\rightarrow 0$, because in this case  the threshold  energy  value
-- the sphaleron  mass -- being proportional to Planck's mass,  goes
to infinity exceeding any  finite  value  of  energy,  so  that  the
transition is always suppressed.


\section{Conclusion}

The main results of our analysis may be summarized  as  follows.  In
the EYM theory, as well as in the pure flat space YM  theory,  there
always exists an opportunity to  pass  over  the  potential  barrier
separating homotopically distinct vacuum sectors.  However,  at  low
energies  all  the  overbarrier  histories  are  suppressed  by  the
destructive interference.  In  the  pure  YM  theory  the  situation
remains the same for any energies. In the EYM theory  on  the  other
hand, when  the  energy  is  large  and  exceeds  the  ground  state
sphaleron mass, the constructive interference occurs  instead.  This
means that the rate of the {\it inclusive} \cite{7}  fermion  number
violating reactions  at  very  high  energies  is  not  small.  Such
reactions lead to the formation of  intermediate  sphaleron  states,
the decay of which will produce a  large  number  of  gravitons  and
gauge bosons \cite{25} as well as extra fermions due to the anomaly.
It has been argued recently by Gibbons and Steif \cite{26}, that the
decay of the BK particles  should  be  accompanied  by  the  fermion
violation.  Our  arguments  show  that,  at   high   energies,   the
probability of BK particles being born and subsequently decaying  is
not small. Notice that the energies available are very large --  the
masses of the EYM sphalerons are of the order  of  $M_{pl}/g$.  Such
processes may arise naturally  within  the  context  of  superstring
theory leading to a ``primordial'' fermion asymmetry. Indeed,  lower
order terms of the expansion of the superstring action in the string
tension, give rise to the coupled Einstein-Yang-Mills-Dilaton (EYMD)
equation of motion. These  equations  possess  classical  solutions,
which resemble in  many  respects  the  BK  solutions  and  coincide
exactly  with  them  in  the  vanishing  dilatonic  coupling   limit
\cite{27}. It is very likely, that these EYMD particles may also  be
interpreted as sphalerons \cite{corn,28}, and our  present  analysis
may be extended to that case.

\section*{Acknowledgments}
I would like to thank Professor Norbert Straumann for discussion and
information about relativistic stars,  Professor  Dmitry  V.Gal'tsov
for discussions and bringing Ref.\cite{charp} into my attention, and
O.Brodbeck for some  conversations.  I  also  would  like  to  thank
M.Heusler, P.Boschung and especially Margaret L.Newens for a careful
reading of the manuscript.

This work was supported by  the  Swiss  National  Science
Foundation.

\appendix
\renewcommand{\thesection}{Appendix \Alph{section}}
\renewcommand{\theequation}{\Alph{section}\arabic{equation}}

\section{The gravitational action}
\setcounter{equation}{0}

In this paper we are interested in Lorenzian  gravitational  action.
In view of the importance of this issue,  we  precisely  specify  in
this Appendix our choice of action and its  relation  to  the  other
possible forms of action.

Our basic requirement for the   action  is  that  it  is  completely
free from second  derivatives  of   the   metric.    Otherwise,  all
the analysis carried in the main  text  becomes  cumbersome.  As  is
known   \cite{ll},   the    second     derivatives    entering   the
Einstein-Hilbert  gravitational   action,   $R\sqrt{-g}$,   may   be
combined to form a total divergence:
\be
R\sqrt{-g}=\Gamma\Gamma\sqrt{-g}  -
\partial_{\mu}(\sqrt{-g}W^{\mu}), \ee
where the two-gamma term is
\be
\Gamma\Gamma=g^{\mu\nu}
(\Gamma^{\beta}_{\mu\alpha}\Gamma^{\alpha}_{\nu\beta} -
\Gamma^{\alpha}_{\mu\nu}\Gamma^{\beta}_{\alpha\beta}), \ee
and
\be
W^{\mu}=g^{\mu\alpha}\Gamma^{\beta}_{\alpha\beta}-
g^{\alpha\beta}\Gamma^{\mu}_{\alpha\beta}=
\frac{1}{\sqrt{-g}}(g^{\mu\alpha}\partial_{\alpha}\sqrt{-g}+
\partial_{\alpha}(\sqrt{-g}g^{\alpha\mu})). \label{w}
\ee
It is natural then  to  choose  for  the  gravitational  action  the
following expression
\be
-\frac{1}{16\pi G}\int (R\sqrt{-g} +
\partial_{\mu}(\sqrt{-g}W^{\mu}))d^{4}x=
-\frac{1}{16\pi    G}\int\Gamma\Gamma\sqrt{-g}\    d^{4}x,
\label{vol}
\ee
because it is manifestly free from all second    derivatives.    Let
the integration in this  formula be performed  over a 4-volume  with
a boundary $\Sigma$ -- we are  working  in  an  asymptotically  flat
spacetime and assume the $\Sigma$ is a distant closed boundary which
is shifted to spatial infinity in the end of the calculations.  Then
(\ref{vol})   may be represented in the following equivalent form
\be
S_{G}=-\frac{1}{16\pi G}\int R \sqrt{-g} d^{4}x -
\frac{1}{16\pi G}\oint_{\Sigma}
n_{\mu}W^{\mu}\sqrt{|^{3}g|}d^{3}x,               \label{ac}
\ee
where the integration in the  second  term  is  performed  over  the
boundary  3-surface,  whose  unit  outward  normal  is  denoted   by
$n^{\mu}$. This expression is our   choice  for  the   action.   The
boundary term in (\ref{ac}) is  non-covariant,  and  quasi-Cartesian
coordinates  are implied  for  calculation  of  it \cite{faddeev} --
we imply that such coordinates exist. To represent this   term in  a
more conventional form, assume that there exists an extension of  the
$n^{\mu}$ vector field into an open neighbourhood of  the  boundary;
then one may write down
\be
\nabla_{\mu}n^{\mu}=\nabla^{\mu}n_{\mu}=
\partial_{\mu}n^{\mu}+\Gamma^{\mu}_{\mu\nu}n^{\nu} =
\partial^{\mu}n_{\mu}-g^{\mu\nu}\Gamma^{\sigma}_{\mu\nu}n_{\sigma},
\ee
which allows us to rewrite the  surface  term  in  Eq.(\ref{ac})  as
follows
\be
-\frac{1}{16\pi G}\oint(n^{\alpha}\Gamma^{\beta}_{\alpha\beta}-
g^{\alpha\beta}\Gamma^{\mu}_{\alpha\beta}n_{\mu})
\sqrt{|^{3}g|}d^{3}x =
\frac{1}{8\pi G}\oint(K+C)d\Sigma,                \label{surf}
\ee
where
\be
K=-\nabla_{\mu}n^{\mu}  \ee
is the trace of the second fundamental form of  the  boundary  which
does not depend on the choice of the extension of $n^{\mu}$, and
\be
C=\frac{1}{2}(\partial_{\mu}n^{\mu}+\partial^{\mu}n_{\mu}).
\label{c}
\ee
is an additional non-covariant quantity. It is easy to see that this
latter quantity does not  involve the {\it  normal}  derivatives  of
the metric. Indeed, representing  $C$  as  $C=\partial_{\mu}n^{\mu}+
\frac{1}{2}g^{\mu\alpha} n^{\sigma} \partial_{\alpha}g_{\mu\sigma}$,
and decomposing the partial derivative  operator,  $\partial_{\mu}$,
into the normal  and  tangential  components  with  respect  to  the
boundary:
\be
\partial_{\mu}= \frac{n_{\mu}}{n^{2}}n^{\alpha}\partial_{\alpha} +
(\delta^{\alpha}_{\mu}-\frac{n_{\mu}n^{\alpha}}{n^{2}})
\partial_{\alpha},
\ee
one can see that the normal derivative of the  metric  in  (\ref{c})
vanishes (the condition $n^{2}=const$ is also   to  be  used).  This
means  that  the  $C$-term  vanishes  on  variation  of  the  action
(\ref{ac}) with  the  boundary condition $\left.  \delta  g_{\mu\nu}
\right|_{\Sigma}=0$, so this term does   not   contribute   to   the
resulting   Einstein equations. Therefore, if  one   is   interested
in   the   action   which  just  reproduces  the  correct   Einstein
equations on variation, then   one   may   drop   the  non-covariant
$C$-term in  Eq.(\ref{surf})  or  replace  it  by  the   some  other
quantity giving no  contribution  on  variation,  for  example,   by
the negative trace of   the   second   fundamental   form   of   the
boundary  imbedded  into  flat space. In this  way  one   comes   to
the  manifestly  covariant   Gibbons   \&  Hawking's   (GH)   action
\cite{14}. It is  clear however,  that   dropping  of   the  surface
$C$-term is  equivalent   to introducing  the   second  (tangential)
derivatives into  the  action,  so   it   is  obvious  that  the  GH
action  is  not completely  free  from   second    derivatives    of
the metric.

Other possible choices of the surface term  also  do  not allow  all
the second derivatives from the action to be excluded; comprehensive
discussion    of    this     topic     can     be      found      in
Ref.\cite{charp}. So, we   should inevitably keep the surface   term
(\ref{surf}) as it stands. The  resulting  first  order  action   is
unique  \cite{charp}.  Thus,  Eq.(\ref{ac})  uniquely  defines   the
gravitational action which is free from the  second  derivatives and
gives correct Einstein equations on variation.

A novel feature arises when an event horizon  is  present.  In  this
case the volume integration  in Eq.(\ref{ac})  should  be  performed
only over the external  with  respect  to  the  horizon region, i.e.
over the part of the 4-volume enclosed by  the   external  boundary,
$\Sigma$, and  the inner boundary -- event  horizon  surface  --  as
well. In obvious notations, the result of such  integration  may  be
represented as follows:
\be
S_{G}=-\frac{1}{16\pi G}\int_{horizon}^{\Sigma}
(\Gamma\Gamma \sqrt{-g}-\partial_{\mu}(\sqrt{-g}W^{\mu})) d^{4}x
-\frac{1}{16\pi G}\oint_{\Sigma}
n_{\mu}W^{\mu}\sqrt{|^{3}g|}d^{3}x.
\ee
Integrating the total divergence, one obtains
\be
S_{G}=-\frac{1}{16\pi G}\int_{(external~region)}
\Gamma\Gamma \sqrt{-g}  d^{4}x -
\frac{1}{16\pi G}\oint_{horizon} n_{\mu}W^{\mu}\sqrt{|^{3}g|}d^{3}x,
\label{hor}
\ee
where the surface integral is to be performed over an inner boundary
in the limit when it tends to the event horizon  surface,  $n^{\mu}$
being the unit outward normal to the boundary. One  may   see   that
the  action  (\ref{ac})  in  this  case  is  not equivalent  to  the
manifestly  first  order  expression   (\ref{vol}).  Thus,  when  an
event horizon is  present,  the  action   (\ref{ac})   does  involve
second   derivatives.   However, these derivatives do not  influence
the variational procedure, as they only  give  the  horizon  surface
contribution to the action, and it is  this  additional  term  which
gives rise  to  the description of the  black   hole   entropy   and
other  related quantities \cite{corn,kallosh}.

Now we perform an explicit computation of the action (\ref{ac})   in
the spherically symmetric case. It is convenient  to  represent  the
metric in the form
\be
ds^{2}=e^{\nu}dt^{2}-e^{\lambda}dr^{2}-r^{2}(d\vartheta^{2}
+sin^{2}\vartheta\ d\varphi^{2}), \label{spherical}\ee
where   $\nu,\lambda$   are   functions  of  $t,r$.   Pass  to   the
quasi-Cartesian    coordinates,    $x^{1}=r\sin\vartheta\cos\varphi,
x^{2}=r\sin\vartheta\sin\varphi,  x^{3}=r\cos\vartheta$. The  metric
then becomes
\be
ds^{2}=e^{\nu}dt^{2}-
(\delta_{ij}+\frac{e^{\lambda}-1}{r^{2}}x_{i}x_{j})dx^{i}dx^{j}
\label{cartesian}
\ee
($x^{i}=x_{i}$). The   quantity   $W^{\mu}$  given  by  Eq.(\ref{w})
reads in these coordinates
\be
W^{\mu}=(W^{0},W^{i})=(\dot{\lambda}exp(-\nu),
\{\frac{2}{r} - (\nu'+\frac{2}{r})exp(-\lambda)\}\frac{x^{i}}{r}).
\label{ww}
\ee
Let us specify the integration domain  in  the  volume  integral  in
Eq.(\ref{ac}) as follows. Let the time, $t$,  vary   from   $-t_{0}$
to  $t_{0}$. We consider the case when an event horizon is  present,
whose  size, $r_{H}$, may  depend  on  time  such   that   $r_{H}(t)
=  r_{H}(-t), r_{H}(\pm t_{0})=0$. Such a choice is specified by the
situation   we  encounter  in  the  main  text,   when   the    time
evolution  of  the metric, whose action we   want   to    calculate,
starts   from   the flat  space   metric   at  $t=-t_{0}$  and  ends
again  at  the flat  space  value  at  $t=t_{0}$ passing, in general
case, through the black hole phase in between. So,  the  integration
domain in this case is: $t\in[-t_{0},t_{0}], r\in[r_{H}(t), R_{0}]$,
the variables  $\vartheta, \varphi$ spanning the spatial two-sphere,
$S^{2}$,  and  the limit  $R_{0} \rightarrow\infty$  being  implied.
The         spacetime          boundary          is         $\Sigma=
\Sigma_{+}\cup\Sigma_{t}\cup\Sigma_{-}$.  The   spacelike  parts  of
the    boundary       are        $\Sigma_{\pm}=\{t=\pm        t_{0},
r\in[0,R_{0}],   \vartheta\in[0,\pi],   \varphi\in[0,2\pi)\}$;   the
corresponding     unit      normals      are      $n_{(\pm)\mu}=(\pm
exp(\nu/2),\vec{0})$;  the        volume        elements         are
$d\Sigma_{\pm}=\sqrt{|^{3}g|}d^{3}x=  exp(\lambda/2)   dr  d\Omega$,
where  $d\Omega=sin\vartheta  d\vartheta  d\varphi$.   The  timelike
part of the boundary  is  $\Sigma_{t}=\{t\in[-t_{0},t_{0}], r=R_{0},
\vartheta\in[0,\pi], \varphi\in[0,2\pi)\}$; the corresponding normal
is   $n_{(t)\mu}=(0,exp(\lambda/2)x^{i}/r)$; the  volume element  is
$d\Sigma_{t}= exp(\nu/2) dt d\Omega$.

Let us calculate first the surface integral in Eq.(\ref{ac}).  Using
Eq.(\ref{ww}) and the definitions introduced, one obtains explicitly
$$
\frac{1}{4\pi}\oint W^{\mu}n_{\mu}d\Sigma =
\left. \iii dr r^{2}\dot{\lambda}exp(\frac{\lambda-\nu}{2})
\right|^{t=t_{0}}_{t=-t_{0}}+
$$
\be
+ \left.\int_{-t_{0}}^{t_{0}} dt\{2r\ exp(\frac{\nu+\lambda}{2}) -
(\nu'r^{2}+2r)exp(\frac{\nu-\lambda}{2})\}\right|_{r=R_{0}}.
\ee
Direct   calculation   of   the   curvature   scalar    gives    the
Einstein-Hilbert term
$$
\frac{1}{4\pi}\int          R\sqrt{-g}d^{4}x  =
\int_{-t_{0}}^{t_{0}}dt \int_{r_{H}(t)}^{R_{0}}dr \{
-\partial_{t}(\dot{\lambda}r^{2}exp(\frac{\lambda-\nu}{2})) +
$$
$$+
\partial_{r}((r^{2}\nu'+2r)exp(\frac{\nu-\lambda}{2})-
2r\ exp(\frac{\nu+\lambda}{2}))+$$\be
+r(\nu'+\lambda')(exp(\frac{\nu+\lambda}{2})-
exp(\frac{\nu-\lambda}{2}))
\}.
\ee
Integrating the total  derivatives,  adding    all    these    terms
together   and   taking   the   limit  $R_{0}\rightarrow\infty$  one
obtains
\be
S_{G}=-\frac{1}{16\pi G}(\int R \sqrt{-g} d^{4}x +\oint_{\Sigma}
n_{\mu}W^{\mu}d\Sigma)= \int_{-t_{0}}^{t_{0}}L_{G}dt, \ee
where
$$
L_{G}=-\frac{1}{4G}\iii
r(\nu'+\lambda')(exp(\frac{\nu+\lambda}{2})-
exp(\frac{\nu-\lambda}{2}))dr -
$$
$$
- \frac{1}{4G}
\{(r^{2}\nu'+2r)exp(\frac{\nu-\lambda}{2})-
2r\ exp(\frac{\nu+\lambda}{2})+
$$
\be
+\left. \dot{r}_{H}(t)\dot{\lambda}r^{2}
exp(\frac{\nu-\lambda}{2})\}\right|_{r=r_{H}(t)}. \label{hren}
\ee
Finally,  introducing  functions  $m(t,r)$, $\sigma(t,r)$   such
that $exp(\nu)=\sigma^{2}(1-2Gm/r)$, $exp(-\lambda)=(1-2Gm/r)$,
$2Gm(t,r_{H}(t)) =r_{H}(t)$,  one  may  see  that  the term involving
time derivatives in (\ref{hren})   vanishes,   and   the   whole
expression  reduces  to  a remarkably simple formula
\be
\left. L_{G}=-\int_{r_{H}(t)}^{\infty}  m\sigma'dr
-\frac{1}{2}\sigma(mr)'
\right|_{r=r_{H}(t)}.  \label{fin}
\ee
The two terms in the right hand side  of  this  equation  relate  to
the volume and horizon terms in Eq.(\ref{hor}) respectively. One may
check  that  for known spherically  symmetric black hole  solutions,
the   expression    (\ref{fin})    yields    a    value     of   the
gravitational  part  of  the  action,   which    gives    rise    to
regular   thermodynamical    parameters    of    a     black    hole
\cite{corn}, \cite{kallosh}.

\section{Spherically symmetric Einstein-Yang-Mills fields.}
\setcounter{equation}{0}

In this Appendix we list  the  full  set  of  the  EYM  equations in
the spherically symmetric case. It is convenient to represent the
gauge-invariant quantites in terms of the variables $\Omega_{0},
\Omega_{1},f$. The non-zero components of the YM field
energy-momentum tensor then are
$$T^{0}_{0}=\frac{1}{2\sigma^{2}}
(\Omega'_{0}-\dot{\Omega}_{1})^{2}       +
\frac{1}{\Delta\sigma^{2}}(\dot{f}^{2}+f^{2}\Omega_{0}^{2})        +
\frac{\Delta}{r^{4}}(f'^{2}+f^{2}\Omega_{1}^{2})                   +
\frac{(f^{2}-1)^{2}}{2r^{4}}, $$
\be T^{r}_{0}=2\frac{\Delta}{r^{4}}(\dot{f}f'+
f^{2}\Omega_{0}\Omega_{1}),                              \label{A2}
\ee
where  $\Delta  =  r^{2}-2mr$.  The  component  $T^{r}_{r}$  can  be
obtained from Eq.(\ref{A2}) by changing the sign  before  the  first
and  the  third  terms on  the  right  hand  side;  other  non-zero
components are  $T^{\vartheta}_{\vartheta}=T^{\varphi}_{\varphi}   =
-(T^{0}_{0}   + T^{r}_{r})$.

The Lagrangian density is
$$g^{2}{\cal L}_{YM}=-\frac{1}{2}trF_{\mu\nu}F^{\mu\nu}\sqrt{-g}=$$
\be
=sin\vartheta\{     \frac{r^{2}}{2\sigma}
(\Omega'_{0}-\dot{\Omega}_{1})^{2}
+
\frac{r^{2}}{\Delta\sigma}(\dot{f}^{2}+f^{2}\Omega_{0}^{2})  -
\frac{\sigma\Delta}{r^{2}}(f'^{2}+f^{2}\Omega_{1}^{2})  -
\sigma\frac{(f^{2}-1)^{2}}{2r^{2}}\} .                 \label{A3}
\ee

The Pontryagin density reads
\be
\frac{1}{16\pi^{2}}trF_{\mu\nu}\tilde{F}^{\mu\nu}\sqrt{-g}=
\frac{sin\vartheta}{8\pi^{2}}\{ (\Omega'_{0}-\dot{\Omega}_{1})
(1-f^{2})+2f(\dot{f}\Omega_{1}-f'\Omega_{0})\}.      \label{A4}
\ee
Non-zero  components  of   the   Chern-Simons   current   given   by
Eq.(\ref{8}) are
$$
K^{0}=\frac{1}{8\pi^{2}\sigma   r^{2}}
\{W_{1}(\k_{1}^{2}+\k_{2}^{2}-1)+\k_{2}\k_{1}'+
(1-\k_{1})\k_{2}'\},
$$
\be
K^{r}=-\frac{1}{8\pi^{2}\sigma r^{2}}
\{W_{0}(\k_{1}^{2}+\k_{2}^{2}-1)+\k_{2}\dot{\k}_{1}+
(1-\k_{1})\dot{\k}_{2}\},                              \label{A5}
\ee
The current is  not, of  course,  gauge  invariant,
but its divergence is, and  this  divergence
is  just  the density (\ref{A4}) (up to the $\sqrt{-g}$ factor).

The non-trivial YM equations read
\be (\frac{r^{2}}{\sigma}(\Omega_{0}'-\dot{\Omega_{1}}))'
=2\frac{r^{2}}{\Delta\sigma}f^{2}\Omega_{0},          \label{A6}
\ee
\be (\frac{r^{2}}{\sigma}(\Omega_{0}'-\dot{\Omega_{1}}))\dot{\ }
=2\frac{\Delta\sigma}{r^{2}}f^{2}\Omega_{1},           \label{A7}
\ee
\be
(\frac{\Delta\sigma}{r^{2}}f')'-(\frac{r^{2}}{\Delta\sigma}
\dot{f})\dot{} =\sigma\frac{f(f^{2}-1)}{r^{2}} +
(\frac{\Delta\sigma}{r^{2}}\Omega_{1}^{2}-
\frac{r^{2}}{\Delta\sigma}\Omega_{0}^{2})f.            \label{A8}
\ee
The independent Einstein equations are
\be m'=r^{2}T^{0}_{0},                                \label{A9}
\ee
\be \dot{m}=-r^{2}T^{r}_{0},                         \label{A10}
\ee
\be
\frac{\sigma'}{\sigma}=\frac{r^{3}}{\Delta}(T^{0}_{0}-
T^{r}_{r}).                                           \label{A11}
\ee

The equations  (\ref{A6})-(\ref{A11})  admit  non-trivial  solutions
discovered by Bartnik and   McKinnon   \cite{9}.   These   solutions
correspond     to    the    static     purely     magnetic     case,
$\Omega_{0}=\Omega_{1}\equiv 0$. The non-trivial  EYM equations then
read
\be
(\frac{\Delta\sigma}{r^{2}}f')'=\sigma\frac{f(f^{2}-1)}{r^{2}},
                                                     \label{A12}
\ee
\be m'=\frac{\Delta}{r^{2}}f'^{2}+\frac{(f^{2}-1)^{2}}{2r^{2}},
                                                     \label{A13}
\ee
\be
\sigma'=\frac{2}{r}f'^{2}\sigma.                     \label{A14}
\ee
The regular asymptotically flat solutions  to  these  equations  are
parameterized  by  an  integer,  $n$; the  corresponding    magnetic
function, $f$, is usually denoted by $w_{n}$. When   $r$  runs  from
zero to infinity, the function $w_{n}(r)$ starts from unit value  at
the  origin  and,  after  $n$  oscillations  around   zero,    tends
asymptotically to $(-1)^{n}$ \cite{9}. The metric functions,  $m(r)$
and  $\sigma(r)$,   monotonically   increase   from   $m(0)=0$    to
$m(\infty)=m_{n}$    and    from     $\sigma(0)=\sigma_{n}<1$     to
$\sigma(\infty) =1$ correspondingly. The asymptotic  value  of   the
mass  function, $m_{n}$, grows  monotonically  with  increasing  $n$
from  the  value $m_{1}=0.828$  to  $m_{\infty}=1$,   the   physical
mass  of  the solutions is  $M_{n}=(\sqrt{4\pi}/g)M_{pl}m_{n}$  with
$M_{pl}$ being Planck's mass.


\section{Variation of the energy functional}
\setcounter{equation}{0}

Consider the barrier height functional defined by Eq.(\ref{49}):
\be
\varepsilon[K(r)]   =    \int_{0}^{\infty}(K'^{2}+\frac{(K^{2}-1)^{2}}
{2r^{2}})exp(-2\int_{r}^{\infty}K'^{2}\frac{dr}{r})dr. \label{B1}
\ee
Let  $K(r)$  be  a  sufficiently  smooth  function  satisfying   the
following conditions
\be
K'(r)=O(r)\ \ \ {\rm as}\ r\rightarrow 0;\ \ \ \
K'(r)=O(1/r^{2})\ \ \ {\rm as}\ r\rightarrow\infty.       \label{B2}
\ee
{\it Define} for convenience two new functions
\be
\sigma(r)=exp\{-2\int_{r}^{\infty}K'^{2}\frac{dr}{r}\} ,\ \ \
m(r)=\frac{1}{\sigma(r)}\int_{0}^{r}(K'^{2} +
\frac{(K^{2}-1)^{2}}{r^{2}})\sigma dr,                  \label{B3}
\ee
they satisfy the following boundary conditions
\be
\sigma(\infty)=1,\ \ \ \sigma(0)\neq 0;\ \ \
 m(\infty)<\infty,\ \ \ m(r)=O(r^{3})\ \ {\rm as}\ r\rightarrow 0,
                                                        \label{B4}
\ee
and also, by definition, the following equations
\be
\sigma'=2\frac{K'^{2}}{r}\sigma,\ \ \ \
 (m\sigma)'=(K'^{2}+\frac{(K^{2}-1)^{2}}{2r^{2}})\sigma. \label{B5}
\ee
Consider small variation
\be K(r)\ \rightarrow\ K(r)+\varphi (r),                \label{B6}
\ee
where
\be \varphi (0)=\varphi(\infty )=0.                     \label{B7}
\ee
To preserve the boundary conditions, the variation must also satisfy
\be \varphi'(r)=O(r)\ \ {\rm as}\ r\rightarrow 0,
\ \ \ \varphi'(r)=O(1/r^{2})\ \ {\rm as}\ r\rightarrow\infty.
                                                          \label{B8}
\ee
Put (\ref{B6}) into (\ref{B1}) and expand the result over
$\varphi$,  one then obtains
\be
\varepsilon[K+\varphi]=\varepsilon[K]+\delta\varepsilon+\delta^{2}
\varepsilon + \ldots ,                                      \label{B9}
\ee
where the first variation is
\be  \delta\varepsilon       =       2\int_{0}^{\infty}\{K'\f'\sigma
+
\frac{K(K^{2}-1)}{r^{2}}\sigma\f     -2I(K'^{2}+\frac{(K^{2}-1)^{2}}
{2r^{2}})\sigma\} dr,                                   \label{B10}
\ee
and the second variation is
\be \delta^{2}\varepsilon = \int_{0}^{\infty}
\{\sigma\f^{2}+\sigma\frac{3K^{2}-1}{r^{2}}\f^{2} -
8I(K'\f'+\frac{K(K^{2}-1)}{r^{2}}\f)\sigma + $$
$$+\ (8I^{2}-2J)(K'^{2}+\frac{(K^{2}-1)^{2}}{2r^{2}})\sigma\}dr,
                                                        \label{B11}
\ee
dots  in  (\ref{B9})   denote   higher   order   terms    and    the
following  new functions have been introduced:
\be I(r)=\int_{r}^{\infty}K'\f'\frac{dr}{r},\ \
J(r)=\int_{r}^{\infty}\f'^{2}\frac{dr}{r}.              \label{B12}
\ee
The boundary conditions (\ref{B2}),(\ref{B8}) imply that
\be I(0)<\infty,\ J(0)<\infty,\ I(\infty)=J(\infty)=0.  \label{B13}
\ee
Consider  the  first  variation   (\ref{B10}).   Using   (\ref{B5}),
   represent
$\delta\varepsilon$ as follows
\be
\delta\varepsilon =2\int_{0}^{\infty}\{K'\f'\sigma+
\frac{K(K^{2}-1)}{r^{2}}\f\sigma - 2I(m\sigma)'\}dr.   \label{B14}
\ee
Integrating by parts one has
\be
\left. \delta\varepsilon=(2K'\sigma\f-4Im\sigma)\right|_{0}^{\infty}+
2\int_{0}^{\infty}\{-(K'\sigma)'\f     +     \sigma\frac{K(K^{2}-1)}
{r^{2}}\f +2m\sigma I'\}dr.                            \label{B15}
\ee
Note, that the boundary terms in  this  expression  vanish.  Finding
$I'$ from (\ref{B12})  and  integrating  by  parts  once  more,  one
finally arrives at
\be
\delta\varepsilon=2\int_{0}^{\infty}\{
-((1-\frac{2m}{r})\sigma
K')'+\frac{K(K^{2}-1)}{r^{2}}\sigma\}\f  dr.\label{B16}
\ee
This  result  agrees  with  Eq.(\ref{50}).  One  can  see  that  the
vanishing  of  the first variation implies the following equation
\be ((1-\frac{2m}{r})\sigma K')'=\sigma\frac{K(K^{2}-1)}{r^{2}}.
                                                       \label{B17}
\ee

Assume now that the first variation vanishes and consider  next  the
second variation. Using (\ref{B5}) one obtains
$$
\delta^{2}\varepsilon=\int_{0}^{\infty}\{\sigma\f'^{
2}+ \sigma\frac{3K^{2}-1}{r^{2}}\f^{2}  -   8I(K'\f'
+  \frac{K(K^{2}-1)}
{r^{2}}\f)\sigma + $$
\be +\ 8I^{2}(m\sigma)'-2J(m\sigma)'\}dr.            \label{B18}
 \ee
Integrating the fourth term in the integrand  by  parts,  and  using
Eq.(\ref{B12}), one obtains
\be
\left.    8\int_{0}^{\infty}I^{2}(m\sigma)'dr    =     8I^{2}m\sigma
\right|_{0}^{\infty}   -    16\int_{0}^{\infty}II'm\sigma    dr    =
16\int_{0}^{\infty}Im\sigma K'\f'\frac{dr}{r} ,      \label{B19}
\ee
where the boundary terms vanish. Combining this result with the
third term in Eq.(\ref{B18}), one has
\be -8\int_{0}^{\infty}I\{ (1-\frac{2m}{r})\sigma        K'\f'    +
\frac{K(K^{2}-1)}{r^{2}}\sigma\f\} dr.               \label{B20}
\ee
Using (\ref{B17}),(\ref{B12}),(\ref{B5}), represent this expression
as follows
$$ -8\int_{0}^{\infty}I((1-\frac{2m}{r})\sigma K'\f)'dr =
\left.  -\  8I(1-\frac{2m}{r})\sigma     K'\f\right|_{0}^{\infty}  +
$$
$$
+\ 8\int_{0}^{\infty}I'(1-\frac{2m}{r})\sigma K'\f dr =
-8\int_{0}^{\infty}(1-\frac{2m}{r})\sigma              K'^{2}\f\f'
\frac{dr}{r}=$$
\be =-4\int_{0}^{\infty}(1-\frac{2m}{r})\sigma'\f\f' dr,\label{B21}
 \ee
where the boundary terms vanish. The fifth term in (\ref{B18})
yields
$$\left.  -2\int_{0}^{\infty}J(m\sigma)'dr  =  -2Jm\sigma\right|_{0}
^{\infty}+  2\int_{0}^{\infty}J'm\sigma  dr  =   -2\int_{0}^{\infty}
m\sigma\f'{2}\frac{dr}{r}= $$
\be = \left. -2m\sigma\frac{\f}{r}\f'\right|_{0}^{\infty} +
2\int_{0}^{\infty}(m\sigma\frac{\f'}{r})'\f dr  = \int_{0}^{\infty}
(\frac{2m}{r}\sigma\f')'\f dr.                      \label{B22}
\ee
The first term in (\ref{B18}) is
\be
\iii \sigma\f'^{2}dr=-\iii(\sigma\f')'dr,            \label{B23}
\ee
where  the  boundary  terms  are  zero.  Consider also the following
expression
\be 0=\iii((1-\frac{2m}{r})\sigma'\f^{2})'dr.        \label{B24}
\ee
Adding  the  equations    (\ref{B21})-(\ref{B24}) and
introducing  also  the tortoise coordinate, $r_{\ast}$,
\be \frac{dr}{dr_{\ast}}= \sigma(1-\frac{2m}{r}),     \label{B25}
\ee
one finally arrives at
\be \delta^{2}\varepsilon = \iii\f(-\frac{d^{2}}{dr_{\ast}^{2}} +V)\f\
dr_{\ast},                                            \label{B26}
\ee
where
\be V=\sigma(1-\frac{2m}{r})\{ 2(\sigma'(1-\frac{2m}{r}))'       +
\frac{3K^{2}-1}{r^{2}}\sigma\},                        \label{B27}
\ee
which  agrees  with  Eq.(\ref{54})  provided  that   Eqs.(\ref{B5}),
(\ref{B17}) are  taken  into account.

\newpage
\begin{center}
{\bf Figure captions}
\end{center}
\vspace{3 cm}

{\bf Fig.1} Typical  shape  of  the  energy-surface  two-dimensional
section $U_{\f}(\lambda,\alpha)$  for  energy  increasing  sphaleron
perturbation modes $\f$. The  surface  forms  a  barrier  separating
distant EYM  vacua  ($\lambda=0,\pi$).  Position  of  the  sphaleron
($\alpha=0,\lambda=\pi/2$) is shown by the vertical arrow.
\vspace{5 mm}

{\bf Fig.2} Plot  of  the  function  $U(\beta,\lambda)$  defined  by
Eq.(\ref{57}).  Vertical  arrow   shows   the   sphaleron   position
$(\beta=1,\lambda=\pi/2)$.  Horizontal  arrows  correspond  to   the
transverse    ($\lambda$-direction)     and     the     longitudinal
($\beta$-direction) sphaleron negative modes.
\vspace{5 mm}

{\bf Fig.3} Behaviour of  the  quantum  phase  $\Phi_{scale}(\beta)$
defined by Eq.(\ref{phase}) with $E=1$. The path passing through the
sphaleron is specified by the value $\beta=1$.

\end{document}